\documentclass[fleqn,usenatbib]{mnras}

\usepackage{newtxtext,newtxmath}
\usepackage[T1]{fontenc}
\usepackage{ae,aecompl}
\usepackage{threeparttable}
\usepackage{graphicx}
\usepackage{subfig}

\usepackage{color, colortbl}
\definecolor{Gray}{gray}{0.9}
\usepackage{diagbox}
\usepackage{booktabs}
\usepackage{multirow}
\usepackage{xfrac}

\usepackage{tipa}
\usepackage{graphicx}	% Including figure files
\usepackage{amsmath}	% Advanced maths commands
\title[Solo IV]{Solo dwarfs IV: Comparing and contrasting satellite and isolated dwarf galaxies in the Local Group} 

\author[Higgs \& McConnachie ]{C.R. Higgs$^{1,2}$\thanks{E-mail: higgs@uvic.ca} and A.W. McConnachie$^{2}$\\
$^{1}$Physics \& Astronomy Department, University of Victoria, 3800 Finnerty Rd, Victoria, B.C., Canada, V8P 5C2\\
$^{2}$NRC Herzberg Astronomy and Astrophysics, 5071 West Saanich
  Road, Victoria, B.C., Canada, V9E 2E7\\
}
\date{Accepted XXX. Received YYY; in original form ZZZ}

\pubyear{2021}

\begin{document}
\label{firstpage}
\pagerange{\pageref{firstpage}--\pageref{lastpage}}

\maketitle

% Abstract of the paper
\begin{abstract}

\noindent We compare and contrast the stellar structures of isolated Local Group dwarf galaxies, as traced by their oldest stellar populations, with the satellite dwarf galaxies of the Milky Way and M\,31. All Local Group dwarfs with $M_v \le -6$ and $\mu_o < 26.5$\,mags\,arcsec$^{-2}$ are considered, taking advantage of measurements from surveys that use similar observations and analysis techniques. For the isolated dwarfs, we use the results from {\it So}litary {\it Lo}cal (Solo) Dwarf Galaxy Survey. 
We begin by confirming that the structural and dynamical properties of the two satellite populations are not obviously statistically different from each other, but we note that there many more satellites around M\,31 than around the Milky Way down to equivalent magnitude and surface brightness limits. We find that dwarfs in close proximity to a massive galaxy generally show more scatter in their Kormendy relations than those in isolation. Specifically, isolated Local Group dwarf galaxies show a tighter trend of half-light radius versus magnitude than the satellite populations, and similar effects are also seen for related parameters. There appears to be a transition in the structural and dynamical properties of the dwarf galaxy population around $\sim 400$\,kpc from the Milky Way and M\,31, such that the smallest, faintest, most circular dwarf galaxies are found closer than this separation. We discuss the impact of selection effects on our analysis, and we argue that our results point to the significance of tidal interactions on the population of systems within approximately 400\,kpc from the MW and M\,31.
\end{abstract}

% Select between one and six entries from the list of approved keywords.
% Don't make up new ones.

%%%%%%%%%%%%%%%%%%%%%%%%%%%%%%%%%%%%%%%%%%%%%%%%%%

\section{Introduction}

Dwarf galaxies are a product of both their external environment and internally-driven evolutionary processes. A massive galaxy in close proximity to a dwarf can have a significant influence on the latter's evolution, affecting properties such as its overall gas content, its star formation rate, and its global stellar structure. 
Within the Local Group, there is a unique opportunity to probe the faintest dwarfs, and to compare those that have evolved close to a large host galaxy to those which exist in isolation. Such an approach has the potential to try to disentangle the roles of nature and nurture at the extreme faint end of the galaxy luminosity function.

The results of this type of analysis to date have provided critical insights into the evolution of dwarf galaxies. There is a well known and long observed morphological transition in the Local Group from dwarf spheroidals (dSphs) to dwarf irregulars (dIrrs) with increasing distance from a massive galaxy \citep{Einasto1974,Spekkens2014, Putman2021}. The transition from star-forming dwarfs with HI gas (dIrrs) to passive dwarfs devoid of gas (dSphs) broadly aligns with the approximate virial radius of the Milky Way (MW) and the Andromeda galaxy (M\,31).  Possible mechanisms at play include ram pressure stripping and tidal stripping, which can remove gas (ram pressure and tides) or stars (tides) from the dwarfs. Interactions may trigger star formation, but gas may also be inhibited from being accreted into satellites (strangulation/starvation), ultimately preventing future star formation. In addition,  the kinematics of the system may be affected on a global scale via a process like tidal stirring (e.g. \citealt{Mayer2006, Kazantzidis2011, Lokas2012}). Here, the interaction induces a bar instability in the disk of the dwarf that redistributes angular momentum and increases the degree of pressure support among the stars. This restructuring is considered necessary to convert rotationally supported systems into pressure supported systems. However, while it is true that gas in Local Group dIrrs is generally seen to rotate (e.g. LITTLE THINGS survey: \citealt{Oh2015}), recent analysis of the stellar kinematics of many low mass dIrrs in the Local Group suggest that the degree of rotation in the old stellar populations is minimal or non-existent (e.g., \citealt{Kirby2014} find only Peg DIG has a clearly rotating stellar component and NGC 6822, IC 1613, UGC 4879, Leo A, Cetus, and Aquarius do not). This perhaps suggests rotation plays a minimal role in stellar systems at the low mass end (e.g., \citealt{Kaufmann2007}). 

With these considerations in mind, we started {\it Solo}, the {\it So}litary {\it Lo}cal) Dwarf Galaxy Survey. The survey is introduced in \citealt[hereafter Paper I]{Higgs2016}. It is designed to target only those nearby galaxies that are further than 300\,kpc from either M\,31 or the Milky Way, and within 3\,Mpc. The 300\,kpc limit was originally motivated by the expected size of the virial radii of these host dark matter halos (e.g., \citealt{klypin2002, Posti2019}). \citealt[hereafter Paper II]{Higgs2021}, analyses the global stellar structure of the dozen isolated Local Group dwarf galaxies within our sample that are visible from the northern hemisphere, focusing on the old (red giant branch) stellar populations.

Our original metric for isolation within the {\it Solo} sample was based upon the present day position of the dwarf galaxy. As such, a necessary caveat to consider is the presence of ``backsplash" galaxies in the isolated sample. These  galaxies are often discussed in the context of galaxy clusters, with a significant fraction of galaxies at 1 - 2 virial radii having had a previous close passage through the center of the cluster (see \citealt{Balogh2000, Gill2005, Smith2015} among others). It is difficult to rule out the same type of scenario for some of the "isolated" Local Group dwarf galaxies (with respect to a previous close passage with the Milky Way or M\,31) without access to the full three-dimension velocities of the dwarfs (however, see \citealt{Buck2019, Teyssier2012}, and \citealt{Blana2020} for careful endeavours to do so). We also need to consider the counterpart: dwarfs that are currently within the virial radius of a massive host but are on their first infall and have not yet had an interaction. 

Most recently, \citealt[hereafter Paper III]{McConnachie2021}, combines our photometry of isolated Local Group galaxies with astrometry from Gaia Data Release 2 (DR2) (\citealt{Gaia2018}) in order to determine the proper motions, and consequently the orbits, for that small subset of {\it Solo} galaxies that have enough bright supergiants that are visible to the Gaia spacecraft (NGC 6822, WLM, IC 1613 and Leo A). For NGC 6822, the Gaia DR2 proper motions appear to favor a scenario where it has never interacted with either of the massive galaxies. For the rest, future Gaia data releases are required in order to provide proper motions with sufficient accuracy to discern their orbital histories.

With these important caveats on the unknown (or only partially known) orbital histories of the systems, this current paper compares the global structural properties of the satellite dwarf galaxies in the Local Group with the isolated Local Group dwarf population. Specifically, we consider their structures as traced via the oldest stellar populations in both sets of galaxies (and as derived for the {\it Solo} sample in Paper II). In Section 2, we introduce the datasets used, including discussion of how we deal with selection effects. In Section \ref{sec:compare}, we compare the structural and derived dynamical parameters of the dwarf populations, including their variation as a function of distance from host galaxy. In Section~4, we discuss our results, and in Section 5 we summarize and conclude.

\section{Sample definition}

The focus of this article is Local Group dwarf galaxies. We consider galaxies to be members of the Local Group if they are within, or at, the zero-velocity surface, as defined by \cite{McConnachie2012}. In practice, this puts all galaxies within a heliocentric distance of 1.4\,Mpc as Local Group members, with the exception of Antlia, Antlia\,B and NGC\,3109 (all members of the nearby NGC\,3109 group). 

Our sample of Local Group satellite galaxies is those systems that are within 300\,kpc from either M\,31 or the Milky Way. Isolated systems are  located more than 300\,kpc from these large galaxies. We refer to these dwarf subsets as the M\,31 satellites, MW satellites and isolated galaxies, and abbreviate to M\,31, MW, and ISO, respectively, in figures. 

In what follows, we only consider galaxies fainter than the Small Magellanic Clouds ($M_v \simeq -16.8 $; i.e., we exclude M33, the LMC and the SMC from this analysis). We additionally require to adopt some faint-end limits to ensure we minimise the effects of varying completeness on our comparisons. For example, the very faintest objects around the MW are more easily found than similar objects around M\,31 or in the outer parts of the Local Group. The completeness of the MW satellite system has been the focus of several recent studies (for example, see \citealt{DrlicaWagner2020, Mao2020, Koposov2015}). 
It is almost certainly the case that the MW system is complete to fainter levels than that of M\,31 or the Local Group as a whole. For M\,31, the most complete census of satellites is from the Pan-Andromeda Archaeological Survey (PAndAS; \citealt{mcconnachie2009}), for example see discussion in \cite{mcconnachie2018}. While a detailed completeness study has not been published, the limiting magnitude of the relevant satellite searches appears to be around $M_V \simeq -6$. For the Local Group, \cite{whiting2007} estimate that their by-eye search for dwarfs in the Palomar Observatory Sky Survey and ESO/Science Research Council survey plates are complete down to a surface brightness of $26 - 27$\,mags\,arcsec$^{-2}$ away from the plane of the Milky Way. This is probably also the appropriate limit for M\,31 satellites outside of the PAndAS footprint. 

Given the considerations above, we only consider dwarfs with an absolute magnitude $M_v \le -6$ and $\mu_o < 26.5$\,mags\,arcsec$^{-2}$. This includes the faintest {\it Solo} dwarf studied by \cite{Higgs2021}, Perseus, at $\mu_o = 26.0 \pm 1.7$\,mags\,arcsec$^{-2}$. We note that we experimented with various completeness cuts, for example considering only $M_v < -8$, and we found the broad conclusions of this paper unchanged (we note that only 2 galaxies in the {\it Solo} sample have $-8 < M_v < -6$). We will discuss remaining issues to do with the completeness of our three samples as they relate to our results at later points in this paper.

For each dwarf subset, we compile a table of their main global positional, structural and dynamical parameters, trying to draw on studies that use homogeneous and consistent observing methods, data reduction procedures and parameter estimations as much as possible:

\begin{itemize}

\item For the isolated dwarfs, we primarily use the recent analysis of structural parameters from Paper II of {\it Solo} for the 12 northern Local Group dwarfs (hereafter referred to as the H21 sample). Here, structural parameters were derived on the basis on the spatial distribution of evolved stellar populations (mostly stars on the red giant branch, RGB), though many of these galaxies also contain significant young stellar populations, which are often brighter and more centrally concentrated. This is a particularly important point, given that the satellite dwarfs are generally dominated by older populations, and makes the comparisons in this current paper less susceptible to being dominated by differences due to only the youngest stars  (see Paper II for full details).
\item For the M\,31 satellites, we primarily use the analysis of 23 dwarfs (all dwarf spheroidals) using PAndAS data by \cite{Martin2016};
\item For the MW satellites, we primarily use the homogeneous analysis from the Megacam Survey of Outer Halo Objects (\citealt{Munoz2018a, Munoz2018b, Marchi-Lasch2019}; hereafter referred to as MSOHO), selecting only those objects that are confirmed dwarf galaxies.  
\end{itemize}

All three of these data sets were observed with CFHT/MegaCam. H21 and PAndAS used the $g, i-$bands and the MSOHO dataset used the $g, r-$bands (supplemented with some additional DECam observations in equivalent filters). Both MSOHO and PAndAS derived parametric fits to the data using a maximum likelihood method. H21 and MSOHO both fit S\'{e}rsic profiles to their dwarfs, whereas the PAndAS subset only used exponential profiles (equivalent to a S\'{e}rsic profile with $n = 1$; the S\'{e}rsic radius is equivalent to a half-light radius). Finally, we note that, for the H21 subset, the central surface brightness and apparent magnitudes were originally determined in the $i-$ band. These values were converted to their corresponding $g-$ band values given the median $(g - i)$ values for the dwarfs. All $g-$band values were transformed to $V-$band using $V = g - 0.098 - 0.544(g - i)$ \citep{Thomas2021}.

\begin{figure}
\centering
	\includegraphics[width=\linewidth]{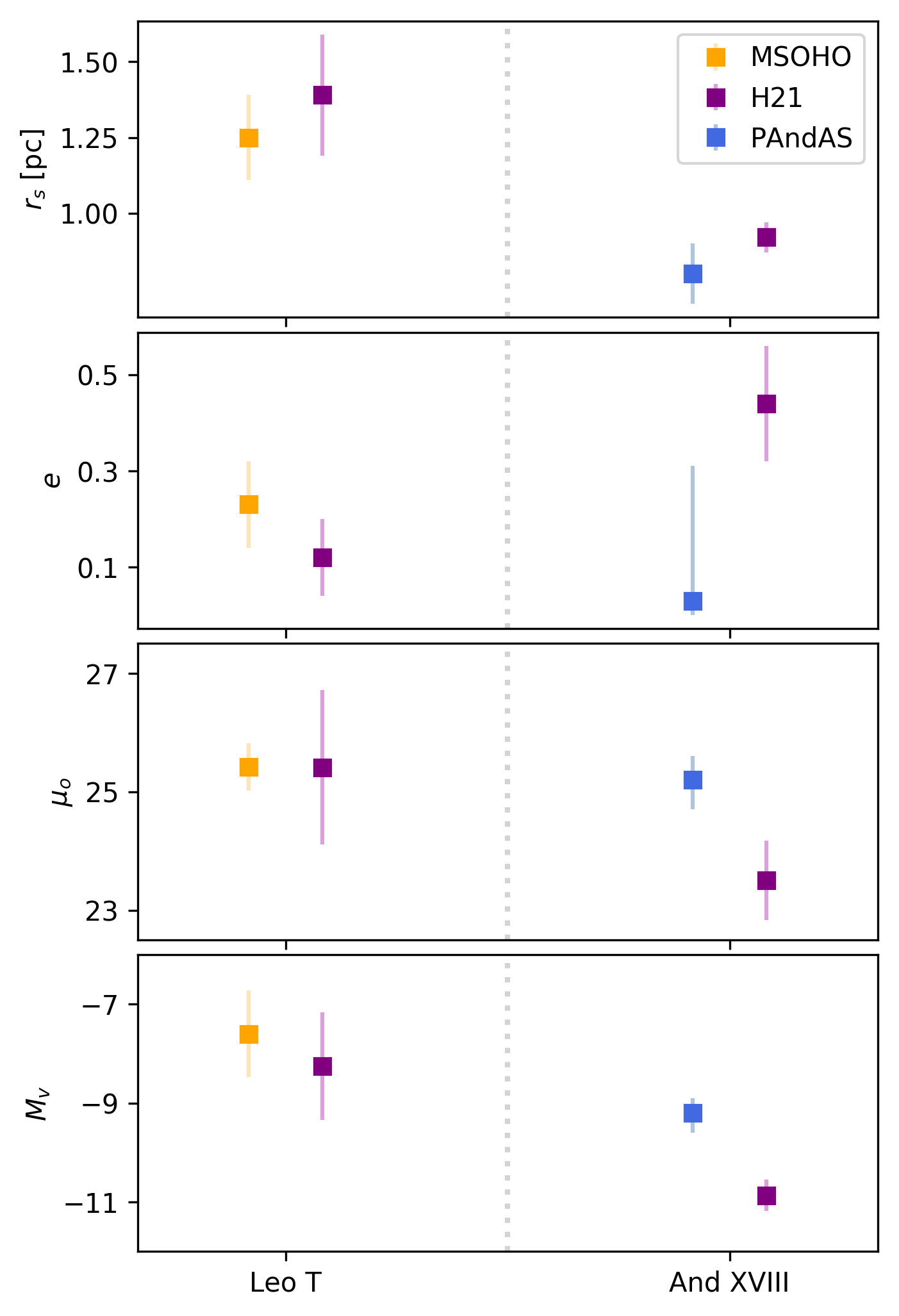}
    \caption{A comparison of measured values for various structural parameters for Leo T and Andromeda XVIII between the studies referenced in the text. } 
    \label{fig:compare}
\end{figure}

It is possible to obtain a crude check on the general consistency or otherwise of the H21, PAndAS and MSOHO studies, since the original papers had a few objects in common. Leo T featured in the MSOHO paper, as well as in H21, and Andromeda XVIII was in the PAndAS paper, as well as H21. Figure~\ref{fig:compare} shows a comparison of various structural parameters between these studies for these two galaxies. We find good agreement between the values derived for Leo T from MSOHO and H21, with measured values lying well within the uncertainties for all parameters.  The half-light radius and ellipiticty of Andromeda XVIII also agree between the PAndAS study and H21; the large uncertainty in the ellipticity from \cite{Martin2016} is a result of much of the galaxy lying on a chip gap in their data. For the surface brightness and apparent magnitude, Paper II measures the galaxy to be considerably brighter than the measurement of \cite{Martin2016}\footnote{The methodology of both Paper II and \cite{Martin2016} is such that surface brightness and magnitude are directly correlated, since the central surface brightness is set by the normalization required for the surface integral of the galaxy profile to equal the galaxy magnitude}. As discussed in Paper II,  this is almost certainly due to the same chip gap, the significance of which is nicely illustrated in Figure 34 of \cite{Martin2016}.

Each of these three primary datasets are supplemented by other works as necessary to ensure all galaxies meeting our selection criteria are included, as well as incorporating velocity dispersion information for each galaxy. Specifically, we use the parameters and references found in the updated \cite{McConnachie2012} catalogue. Tables~\ref{tab:iso}, \ref{tab:mw} and \ref{tab:M31} list the parameters and sources for the isolated, MW and M\,31 dwarf subsets respectively. More specifically, the columns are:
\begin{itemize}
    \item positions: Right ascension (RA) and declination (Dec.)
    \item distance moduli: $(m-M)$
    \item ellipticity: $e$
    \item central surface brightness in the V-band: $\mu_{vo}$
    \item stellar velocity dispersion: $\sigma_s$
    \item absolute magnitude in the V-band: $M_v$
    \item Sersic index: $n$ (where known) 
    \item half-light radii: $r_s$
    \item dynamical mass: M\textsubscript{dyn} (computed using Eqn. 1, given later)
    \item dynamical mass to light ratio: $M/L$ (computed using the dynamical masses and observed luminosity)
    \item minimum distance to a large galaxy (MW or M\,31): D\textsubscript{min} (computed using positions listed here)
\end{itemize}

\begin{table*}
\begin{threeparttable}
\centering
\resizebox{\textwidth}{!}{\begin{tabular}{l||cccccccccccccccc}
  Name & RA & Dec. & $(m-M)$ & $e$ & $\mu_{vo}$ & $\sigma_s$ & M\textsubscript{v} & $n$ & $r_s$ & Ref. & M\textsubscript{dyn} & $M/L$ & D\textsubscript{min}\\
     &  &  &  &  &  & [km s$^{-1}$] &  &  & [pc] & & [$10^6$ M$_{\odot}$] & [M$_{\odot}$ L$_{\odot}^{-1}$] & [kpc]\\
\hline
  And XXVIII           & $22^h 32^m 41.2^s$ & $+31^{\circ}12'58.0"$  & 24.44$\pm$0.04 & 0.42$\pm$0.06 & 25.5$\pm$0.9 & 6.6$^{+2.9}_{-2.1}$ & -9.1$\pm$0.5 & 0.84$\pm$0.13 & 316$\pm$22 & {\it 1,3} & 12.6$^{+11}_{-8}$ & 61$^{+92}_{-43}$ & 374 \\
  And XXXIII (Perseus) & $03^h 01^m 23.6^s$ & $+40^{\circ}59'18.0"$  & 24.18$^{+0.11}_{-0.09}$ & 0.04$\pm$0.08 & 26.0$\pm$1.7 & - & -8.9$\pm$1.8 & 0.79$\pm$0.63 & 276$\pm$41 & {\it 1} & - & - & 341\\
  IC 1613 (DDO 8)     & $01^h 04^m 47.8^s$ & $+02^{\circ}07'4.0"$   & 24.18$\pm$0.06 & 0.20$\pm$0.05 & 23.3$\pm$0.4 & 10.8$^{+1.0}_{-0.9}$ & -15.72$\pm$0.09 & 0.43$\pm$0.02 & 1511$\pm$42 & {\it 1,5} & 164$^{+31}_{-28}$ & 2.0$\pm$0.4 & 503\\
  Phoenix              & $01^h 51^m 06.3^s$ & $-44^{\circ}26'41.0"$  & 23.06$\pm$0.12 & 0.30$\pm$0.03 & 25.8 & 9.3$\pm$0.7 & -9.89$\pm$0.42 & - & 274$\pm$18 & {\it 10,13,14} & 22$^{+4}_{-3}$ & 57$^{+30}_{-19}$ & 409 \\
  NGC 6822 (DDO 209)   & $19^h 44^m 56.6^s$ & $-14^{\circ}47'21.0"$  & 23.78$\pm$0.05 & 0.28$\pm$0.15 & 22.5$\pm$0.04 & 23.2$\pm$1.2 & -16.47$\pm$0.08 & 0.73$\pm$0.02 & 1980$\pm$45 & {\it 1,5} & 992$^{+105}_{-101}$ & 6.0$^{+0.8}_{-0.7}$ & 563\\
  Cetus                & $00^h 26^m 11.0^s$ & $-11^{\circ}02'40.0"$  & 24.39$\pm$0.07 & 0.33$\pm$0.06 & 25.0$\pm$0.2 & 8.3$\pm$1.0 & -11.29$\pm$0.21 & - & 703$\pm$32 & {\it 5,6,7,15} & 45$^{+12}_{-10}$ & 32$^{+11}_{-9}$ & 679\\
  Peg DIG (DDO 216)    & $23^h 28^m 36.3^s$ & $+14^{\circ}44'35.0"$  & 24.77$\pm$0.04 & 0.56$\pm$0.05 & 23.3$\pm$0.5 & 12.3$^{+1.2}_{-1.1}$ & -13.6$\pm$0.1 & 0.77$\pm$0.03 & 995$\pm$25 & {\it 1,5} & 140$^{+28}_{-25}$ & 12$^{+3}_{-2}$ & 463\\
  Leo T                & $09^h 34^m 53.4^s$ & $+17^{\circ}03'5.0"$   & 23.08$\pm$0.08 & 0.12$\pm$0.08 & 25.4$\pm$1.3 & 7.5$\pm$1.6 & -8.3$\pm$1.1 & 0.86$\pm$0.36 & 167$\pm$25 & {\it 1,4} & 9$^{+4}_{-3}$ & 100$^{+191}_{-67}$ & 418\\
  WLM (DDO 221)        & $00^h 01^m 58.2^s$ & $-15^{\circ}27'39.0"$  & 24.85$\pm$0.05 & 0.54$\pm$0.06 & 22.7$\pm$0.5 & 17.5$\pm$2.0 & -14.52$\pm$0.15 & 0.77$\pm$0.04 & 1114$\pm$44 & {\it 1,12} & 300$^{+37}_{-36}$ & 10$^{+1.5}_{-0.9}$ & 835\\
  And XVIII            & $00^h 02^m 14.5^s$ & $+45^{\circ}05'20.0"$  & 25.36$\pm$0.08 & 0.44$\pm$0.12 & 23.5$\pm$0.7 & 9.7$\pm$2.3 & -10.9$\pm$0.3 & 0.95$\pm$0.10 & 316$\pm$22 & {\it 1,11} & 27$^{+15}_{-12}$ & 28$^{+20}_{-13}$ & 424\\
  Leo A (DDO 69)      & $09^h 59^m 26.5^s$ & $+30^{\circ}44'47.0"$  & 24.28$\pm$0.05 & 0.42$\pm$0.05 & 24.0$\pm$0.6 & 9.0$^{+0.8}_{-0.6}$ & -11.7$\pm$0.3 & 0.72$\pm$0.72 & 479$\pm$21 & {\it 1,2} & 36$^{+6}_{-5}$ & 18$^{+6}_{-5}$ & 722\\
  Aquarius (DDO 210)   & $20^h 46^m 51.8^s$ & $-12^{\circ}50'53.0"$  & 24.97$\pm$0.05 & 0.53$\pm$0.05 & 24.7$\pm$0.9 & 7.8$^{+1.8}_{-1.1}$ & -10.92$\pm$0.18 & 0.61$\pm$0.05 & 468$\pm$26 & {\it 1,2} & 26$^{+11}_{-9}$ & 26$^{+15}_{-11}$ & 981 \\%463\\
  Tucana               & $22^h 41^m 49.6^s$ & $-64^{\circ}25'10.0"$  & 24.74$\pm$0.12 & 0.48$\pm$0.03 & 25.0$\pm$0.1 & 15.8$^{+4.1}_{-3.1}$ & -9.54$\pm$0.23 & - & 284$\pm$54 & {\it 8,9,16} & 64$^{+37}_{-27}$ & 232$^{+168}_{-113}$ & 883\\
  Sag DIG              & $19^h 29^m 59.0^s$ & $-17^{\circ}40'51.0"$  & 25.39$\pm$0.08 & 0.56$\pm$0.18 & 24.1$\pm$0.7 & 9.4$^{+1.5}_{-1.1}$ & -11.4$\pm$0.2 & 0.75$\pm$0.07 & 497$\pm$34 & {\it 1,2} & 40$^{+12}_{-11}$ &26$^{+12}_{-9}$ & 1190\\
  UGC 4879 (VV 124)    & $09^h 16^m 02.2^s$ & $+52^{\circ}50'24.0"$  & 25.42$\pm$0.06 & 0.43$\pm$0.06 & 22.7$\pm$0.9 & 9.6$^{+1.3}_{-1.2}$ & -11.7$\pm$0.6 & 1.28$\pm$0.17 & 400$\pm$35 & {\it 1,5} & 34$^{+10}_{-9}$ & 17$^{+13}_{-7}$  & 1219\\

\end{tabular}}
\caption{Relevant observational parameters, and derived quantities, for the isolated dwarf dataset.}
    \label{tab:iso}
    \begin{tablenotes}
    \scriptsize
    
        \item[] {\it 1} - H21 (\citealt{Higgs2021}), {\it 2} - \cite{Kirby2017} %vel disp
         {\it 3} - \cite{Collins2013}, %vel disp
         {\it 4} - \cite{Simon2007}, %vel disp
        {\it 5} - \cite{Kirby2014}, %vel disp
        {\it 6} - \cite{McConnachie2005},
         {\it 7} - \cite{McConnachieIrwin2006},
       \item[]  {\it 8} - \cite{Fraternali2009},
         {\it 9} - \cite{Saviane1996},
         {\it 10} - \cite{Battaglia2012},
         {\it 11} - \cite{Tollerud2012},
         {\it 12} - \cite{Leaman2012},
         {\it 13} - \cite{Kacharov2017},
         {\it 14} - \cite{MartinezDelgado1999a},
        \item[]  {\it 15} - \cite{Lewis2007},
         {\it 16} - \cite{Bernard2009}

    \end{tablenotes}
\end{threeparttable}
\end{table*}
  
\begin{table*}
\begin{threeparttable}
\centering
\resizebox{\textwidth}{!}{\begin{tabular}{l||ccccccccccccc}
  Name & RA & Dec. & $(m-M)$ & $e$ & $\mu_{vo}$ & $\sigma_s$ & M\textsubscript{v} & $n$ & $r_s$ & Ref. & M\textsubscript{dyn} & $M/L$ & D\textsubscript{min}\\
     &  &  &   &  &  & [km s$^{-1}$] &  &  & [pc] & & [$10^6$ M$_{\odot}$] & [M$_{\odot}$ L$_{\odot}^{-1}$] & [kpc]  \\
\hline
  Sag Sph    & $18^h 55^m 19.5^s$ & $-30^{\circ}32'43.0"$ & 17.10$\pm$0.15 & 0.64$\pm$0.02 & 25.2$\pm$0.3 & 11.4$\pm$0.7 & -13.5$\pm$0.3 & - & 2600$\pm$200 & {\it 12,13,14,15} & 316$^{+48}_{-43}$ &  29$^{+12}_{-9}$ & 19\\
  Draco (DDO 208) & $17^h 20^m 12.4^s$ & $+57^{\circ}54'55.0"$ & 19.40$\pm$0.17  & 0.30$\pm$0.01 & 25.12$\pm$0.07 & 9.1$\pm$1.2$^{}$ & -8.71$\pm$0.17 & 0.96$\pm$0.02 & 219$\pm$18  & {\it 1,2,3} & 17$^{+5}_{-4}$ & 128$^{+45}_{-36}$ & 76\\
  Ursa Minor (DDO 199) & $15^h 09^m 08.5^s$ & $+67^{\circ}13'21.0"$ & 19.40$\pm$0.10 & 0.55$\pm$0.01 & 25.77$\pm$0.08 & 9.5$\pm$1.2$^{}$ & -9.03$\pm$0.12 & 0.82$\pm$0.01 & 382$\pm$18 & {\it 1,3,4,5} & 32$^{+9}_{-8}$ & 183$^{+56}_{-47}$ & 78 \\
  Sculptor   & $01^h 0^m 09.4^s$ & $-33^{\circ}42'33.0"$ & 19.67$\pm$0.14 & 0.37$\pm$0.01 & 23.29$\pm$0.15 & 9.2$\pm$0.2$^{}$ & -10.82$\pm$0.14 & 1.16$\pm$0.01 & 308$\pm$21 & {\it 1,6,10} & 26$^{+10}_{-8}$ & 27$\pm$9 & 86 \\
  Carina     & $06^h 41^m 36.7^s$ & $-50^{\circ}57'58.0"$ & 20.11$\pm$0.13 & 0.37$\pm$0.01 & 25.35$\pm$0.07 & 6.7$\pm$0.3$^{}$ & -9.43$\pm$0.15 & 0.94$\pm$0.01 & 349$\pm$22 & {\it 1,7,10} & 14$^{+6}_{-5}$ & 55$^{+25}_{-19}$ & 107 \\
  Fornax     & $02^h 39^m 59.3^s$ & $-34^{\circ}26'57.0"$ & 20.84$\pm$0.18 & 0.28$\pm$0.01 & 23.59$\pm$0.16 & 11.8$\pm$0.2$^{}$ & -13.46$\pm$0.24 & 0.71$\pm$0.01 & 788$\pm$69 & {\it 1,7,10} & 100$^{+19}_{-16}$ & 10$^{+3}_{-2}$ & 149\\
  Leo II (DDO 93) & $11^h 13^m 28.8^s$ & $+22^{\circ}09'06.0"$ & 21.84$\pm$0.13 & 0.07$\pm$0.02 & 24.24$\pm$0.07 & 6.6$\pm$0.7$^{}$ & -9.74$\pm$0.16 & 0.71$\pm$0.02 & 168$\pm$11 & {\it 1,3,8} & 6.8$^{+1.6}_{-1.4}$ & 20$^{+6}_{-5}$ & 236\\
  Leo I (DDO 74) & $10^h 08^m 28.1^s$ & $+12^{\circ}18'23.0"$ & 22.02$\pm$0.13 & 0.30$\pm$0.01 & 22.61$\pm$0.30 & 9.2$\pm$0.4$^{}$ & -11.78$\pm$0.33 & 0.77$\pm$0.02 & 243$\pm$15 & {\it 1,9,11} & 19$\pm$2 & 9$^{+3}_{-2}$ & 257\\

\end{tabular}}
    \caption{Relevant observational parameters, and derived quantities, for the Milky Way satellite dataset.}
    \label{tab:mw}
     \begin{tablenotes}
     \scriptsize
        \item[] {\it 1} - MSOHO (\citealt{Munoz2018b}),{\it 2} - \cite{Bonanos2004}, {\it 3} - \cite{Walker2007}, {\it 4} - \cite{Walker2009b}, {\it 5} - \cite{Carrera2002}, {\it 6} - \cite{Pietrzynski2008}, {\it 7} - \cite{Pietrzynski2009}, 
        \item[] {\it 8} - \cite{Bellazzini2005}, {\it 9} - \cite{Bellazzini2004}, {\it 10} - \cite{Walker2009a}, {\it 11} - \cite{Mateo2008}, {\it 12} - \cite{Monaco2004}, {\it 13} - \cite{Ibata1997}, {\it 14} - \cite{Mateo1998b}, 
         \item[] {\it 15} - \cite{Majewski2003}

    \end{tablenotes}
\end{threeparttable}
\end{table*}

%\begin{minipage}{\linewidth}
\begin{table*}
\begin{threeparttable}
    \centering
\resizebox{\textwidth}{!}{\begin{tabular}{l||ccccccccccccccc}
  Name & RA & Dec. & $(m-M)$ & $e$ & $\mu_{vo}$ & $\sigma_s$ & M\textsubscript{v} & $n$ & $r_s$ & Ref. & M\textsubscript{dyn} & $M/L$ & D\textsubscript{min}\\
     &  &  &   &  &  & [km s$^{-1}$] &  &  & [pc] & & [$10^6$ M$_{\odot}$] & [M$_{\odot}$ L$_{\odot}^{-1}$]  & [kpc]\\
\hline
  M 32      & $00^h42^m41.8^s$ & $+40^{\circ}51'55.0"$ & 24.53$\pm$0.21 & 0.25$\pm$0.02 & 11.1 & 92.0$\pm$5.0\tnote{} & -16.4$\pm$0.2 & - & 110$\pm$16 & {\it 7,15,16,17} & 861$^{+164}_{-147}$ & 5.4$^{+1.7}_{-1.3}$ & 26 \\
  NGC 205   & $00^h40^m22.1^s$ & $+41^{\circ}41'07.0"$ & 24.58$\pm$0.07 & 0.43$\pm$0.10 & 19.2 & 35.0$\pm$5.0\tnote{} & -16.48$\pm$0.12 & - & 590 $\pm$ 16 & {\it 3,15,16} & 667$^{+209}_{-178}$ & 4.0$^{+1.4}_{-1.1}$ & 45 \\
  And I     & $00^h45^m39.8^s$ & $+38^{\circ}02'28.0"$ & 24.36$\pm$0.07 & 0.28$\pm$0.3 & 25.4$\pm$0.2 & 10.2$\pm$1.9\tnote{} & -11.2$\pm$0.2 & - & 815$\pm$40 & {\it 1,3,4} & 79$^{+32}_{-27}$ & 61$^{+28}_{22}$ & 56\\
  And XVII  & $00^h37^m07.0^s$ & $+44^{\circ}19'20.0"$ & 24.31$^{+0.11}_{-0.08}$ & 0.50$\pm$0.10 & 26.4$^{+0.4}_{-0.3}$ & 2.9$^{+2.2}_{-1.9}$\tnote{} & -7.8$\pm$0.3 & - & 285$^{+55}_{-45}$ & {\it 1,2,8} & 2.2$^{+4.4}_{-1.9}$ & 39$^{+84}_{-34}$ & 67\\
  And III   & $00^h35^m33.8^s$ & $+36^{\circ}29'52.0"$ & 24.37$\pm$0.07 & 0.59$\pm$0.04 & 25.1$\pm$0.3 & 9.3$\pm$1.4\tnote{} & -9.5$\pm$0.3 & - & 405$\pm$35 & {\it 1,3,4} & 33$^{+11}_{-9}$ & 119$^{+59}_{-41}$ & 73 \\
  And V     & $01^h10^m17.1^s$ & $+47^{\circ}37'41.0"$ & 24.44$\pm$0.08 & 0.26$^{+0.09}_{-0.07}$ & 25.6$\pm$0.3 & 11.5$^{+5.4}_{-4.4}$\tnote{} & -9.3$\pm$0.2 & - & 345$\pm$40 & {\it 1,2,3} & 47$^{+20}_{-16}$ & 211$^{+112}_{-83}$ & 109\\
  And XX    & $00^h07^m30.7^s$ & $+35^{\circ}07'56.0"$ & 24.35$^{+0.12}_{-0.16}$ & 0.11$^{+0.41}_{-0.11}$ & 25.8$\pm$0.7 & 7.1$^{+3.9}_{-2.5}$\tnote{} & -6.4$^{+0.5}_{-0.4}$ & - & 90$^{+35}_{-20}$  & {\it 1,2,3,8} & 4$^{+5}_{-3}$ & 242$^{+465}_{-197}$ & 128\\
  And XXXII (Cassiopeia III) & $00^h35^m59.4^s$ & $+51^{\circ}33'35.0"$ & 24.45$\pm$0.14 & 0.50$\pm$0.09 & 26.4$\pm$0.8 & 8.4$\pm$0.6\tnote{} & -12.3$\pm$0.7 & - & 1468$^{+288}_{-244}$ & {\it 6,10}\tnote{3} & 142$^{+40}_{-34}$ & 41$^{+42}_{-21}$ & 140\\
  NGC 147 (DDO 3)   & $00^h33^m12.1^s$ & $+48^{\circ}30'32.0"$ & 24.15$\pm$0.09 & 0.41$\pm$0.02 & 21.2 & 16.0$\pm$1.0\tnote{} & -14.65$\pm$0.13 & - & 623 & {\it 3,15,18} & 148$^{+19}_{-18}$ & 4.8$^{+0.9}_{-0.8}$ & 140\\
  And XXX   & $00^h36^m34.9^s$ & $+49^{\circ}38'48.0"$ & 24.17$^{+0.10}_{-0.26}$ & 0.43$^{+0.10}_{-0.12}$ & 26.1$\pm$0.3 & 11.8$^{+7.7}_{-4.7}$\tnote{} & -8.0$^{+0.4}_{-0.3}$ & - & 270$\pm$50 & {\it 1,2,8} & 33$^{+47}_{-26}$ & 488$^{+957}_{-411}$ & 145 \\
  And XIV   & $00^h51^m35.0^s$ & $+29^{\circ}41'49.0"$ & 24.50$^{+0.06}_{-0.56}$ & 0.17$^{+0.16}_{-0.17}$ & 26.3$\pm$0.3 & 5.3$\pm$1.0\tnote{} & -8.5$^{+0.4}_{-0.3}$ & - & 265 & {\it 1,4,8} & 7$^{+3}_{-2}$ & 64 $^{+36}_{-25}$ & 161\\
  And XV    & $01^h14^m18.7^s$ & $+38^{\circ}07'03.0"$ & 23.98$^{+0.26}_{-0.12}$ & 0.24$\pm$0.10 & 26.1$\pm$0.3 & 4.0$\pm$1.4\tnote{} & -8.0$^{+0.3}_{-0.4}$ & - & 230$^{+35}_{-25}$ & {\it 1,4,8} & 3$^{+3}_{-2}$ & 48$^{+53}_{-29}$ & 176\\
  And II    & $01^h16^m29.8^s$ & $+33^{\circ}25'09.0"$ & 24.07$\pm$0.06 & 0.16$\pm$0.02 & 25.6$\pm$0.2 & 7.8$\pm$1.1\tnote{} & -11.6$\pm$0.2 & - & 965$\pm$45 & {\it 1,3,9} & 55$^{+17}_{-14}$ & 29$^{+11}_{-9}$ & 181\\
  NGC 185   & $00^h38^m58.0^s$ & $+48^{\circ}20'15.0"$ & 23.95$\pm$0.09 & 0.15$\pm$0.01 & 20.8 & 24.0$\pm$1.0\tnote{} & -14.75$\pm$0.13 & - & 457 & {\it 3,15,18} & 245$\pm$20 & 7.2$^{+1.1}_{-1.0}$ & 184 \\
  And XXIX  & $23^h58^m55.6^s$ & $+30^{\circ}45'20.0"$ & 24.32$\pm$0.22 & 0.35$\pm$0.06 & 26.4$\pm$0.9 & 5.7$\pm$1.2\tnote{} & -8.3$\pm$0.5 & - & 362$^{+57}_{-55}$ & {\it 13,14} & 11$^{+6}_{-4}$ & 116$^{+93}_{-54
  }$ & 187\\
  And VII (Cassiopeia)   & $23^h26^m31.7^s$ & $+50^{\circ}40'33.0"$ & 24.41$\pm$0.10 & 0.13$\pm$0.04 & 23.2$\pm$0.2 & 13.0$\pm$1.0\tnote{} & -13.2$\pm$0.3 & - & 775 $\pm$43 & {\it 3,4,11,19} & 121$^{+20}_{-19}$ & 15$^{+6}_{-4}$ & 218\\
  IC 10     & $00^h20^m17.3^s$ & $+59^{\circ}18'14.0"$ & 24.50$\pm$0.12 & 0.19$\pm$0.02 & 24.6$\pm$0.2 & 35.5$\pm$16.6\tnote{} & -15.0$\pm$0.2 & - & 612 & {\it 5,15,20,21,22} & 716$^{+796}_{-511}$ & 17$^{+20}_{-12}$ & 252\\
  And XXXI (Lacerta I)  & $22^h58^m16.3^s$ & $+41^{\circ}17'28.0"$ & 24.40$\pm$0.12 & 0.43$\pm$0.07 & 25.8$\pm$0.8 & 10.3$\pm$0.9\tnote{} & -11.7$\pm$0.7 & - & 927$^{+102}_{-121}$ & {\it 6,10} & 60$^{+10}_{-11}$ & 29$^{+28}_{-14}$ & 262\\
  LGS 3 (Pisces)    & $01^h03^m55.0^s$ & $+21^{\circ}53'06.0"$ & 24.43$\pm$0.07 & 0.2$\pm$0.1 & 24.8$\pm$0.1 & 7.9$^{+5.3}_{-2.9}$\tnote{} & -10.13$\pm$0.13 & - & 470  $\pm$ 47 & {\it 3,12,23} & 27$^{+36}_{-21}$ & 58$^{+99}_{-49}$ & 268\\
  And VI (Peg dSph)    & $23^h51^m46.3^s$ & $+24^{\circ}34'57.0"$ & 24.47$\pm$0.07 & 0.41$\pm$0.03 & 24.1$\pm$0.2 & 12.4$^{+1.5}_{-1.3}$\tnote{} & -11.5$\pm$0.2 & - & 524 $\pm$49 & {\it 2,3,11} & 74$^{+20}_{-17}$ & 45$^{+17}_{-13}$ & 268\\
  And XVI   & $00^h59^m29.8^s$ & $+32^{\circ}22'36.0"$ & 23.39$^{+0.19}_{-0.14}$ & 0.29$\pm$0.08 & 25.5$\pm$0.3 & 3.8$\pm$2.9\tnote{} & -7.3$\pm$0.3 & - & 130 $^{+30}_{-15}$& {\it 1,4,8} & 1.7$^{+3.7}_{-1.5}$ & 48$^{+105}_{-42}$ & 320\\

\end{tabular}}
    \caption{Relevant observational parameters, and derived quantities, for the M\,31 satellite dataset.}
    \label{tab:M31}
    \begin{tablenotes}
    \scriptsize
         \item[] {\it 1} - PAndAS (\citealt{Martin2016}),
         {\it 2} - \cite{Collins2013}, %Velocity dispersions
         {\it 3} - \cite{McConnachie2005},%distances
         {\it 4} - \cite{Tollerud2012}, %Velocity dispersions
        \item[]{\it 5}  \cite{Ho2009} (based on integrated light rather than resolved stars unlike the rest of the measurements),
        {\it 6} - \cite{Martin2014},
        {\it 7} - \cite{Fiorentino2010},
        {\it 8} - \cite{Conn2012},
        {\it 9} - \cite{Ho2012},
        \item[]{\it 10} - \cite{Martin2013a},
        {\it 11} - \cite{McConnachieIrwin2006},
        {\it 12} - \cite{Lee1995},
        {\it 13} - \cite{Tollerud2013},
        {\it 14} - \cite{Bell2011},
        {\it 15} - \cite{Devaucouleurs1991},
        {\it 16} - \cite{Choi2002},
        \item[]{\it 17} - \cite{Grillmair1996},
        {\it 18} - \cite{Geha2010},
        {\it 19} - \cite{Kalirai2010},
        {\it 20} - \cite{Huchra1999},
        {\it 21} - \cite{Sanna2010},
        {\it 22} - \cite{Tikhonov2009},
        {\it 23} - \cite{Cook1999}

    \end{tablenotes}
\end{threeparttable}
\end{table*}

\section{Comparing and contrasting the dwarf galaxy populations}\label{sec:compare}

Here, we compare the distributions of the various parameters listed in Tables~\ref{tab:iso}, \ref{tab:mw} and \ref{tab:M31} for each of the dwarf populations. We hold off on discussion of the relevant figures until Section~\ref{sec:discussion}.

\subsection{Structural comparisons}\label{sec:structure}

\begin{figure*}%
    \centering
    \subfloat[ S\'{e}rsic radius ($r_s$,pc)] {{\includegraphics[width=0.45\linewidth]{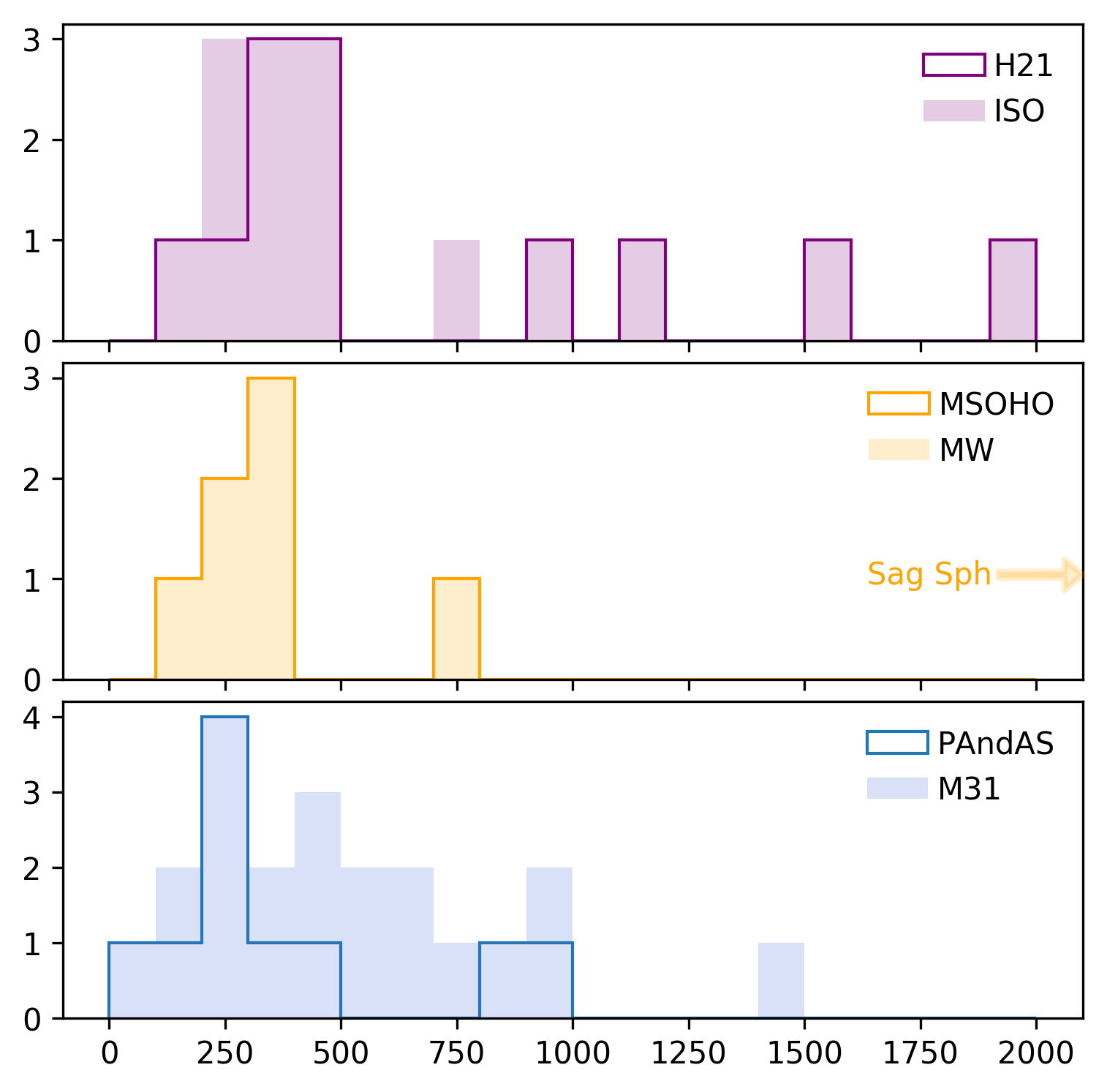} }}%
    \subfloat[Ellipticity ($e$)]{{\includegraphics[width=0.45\linewidth]{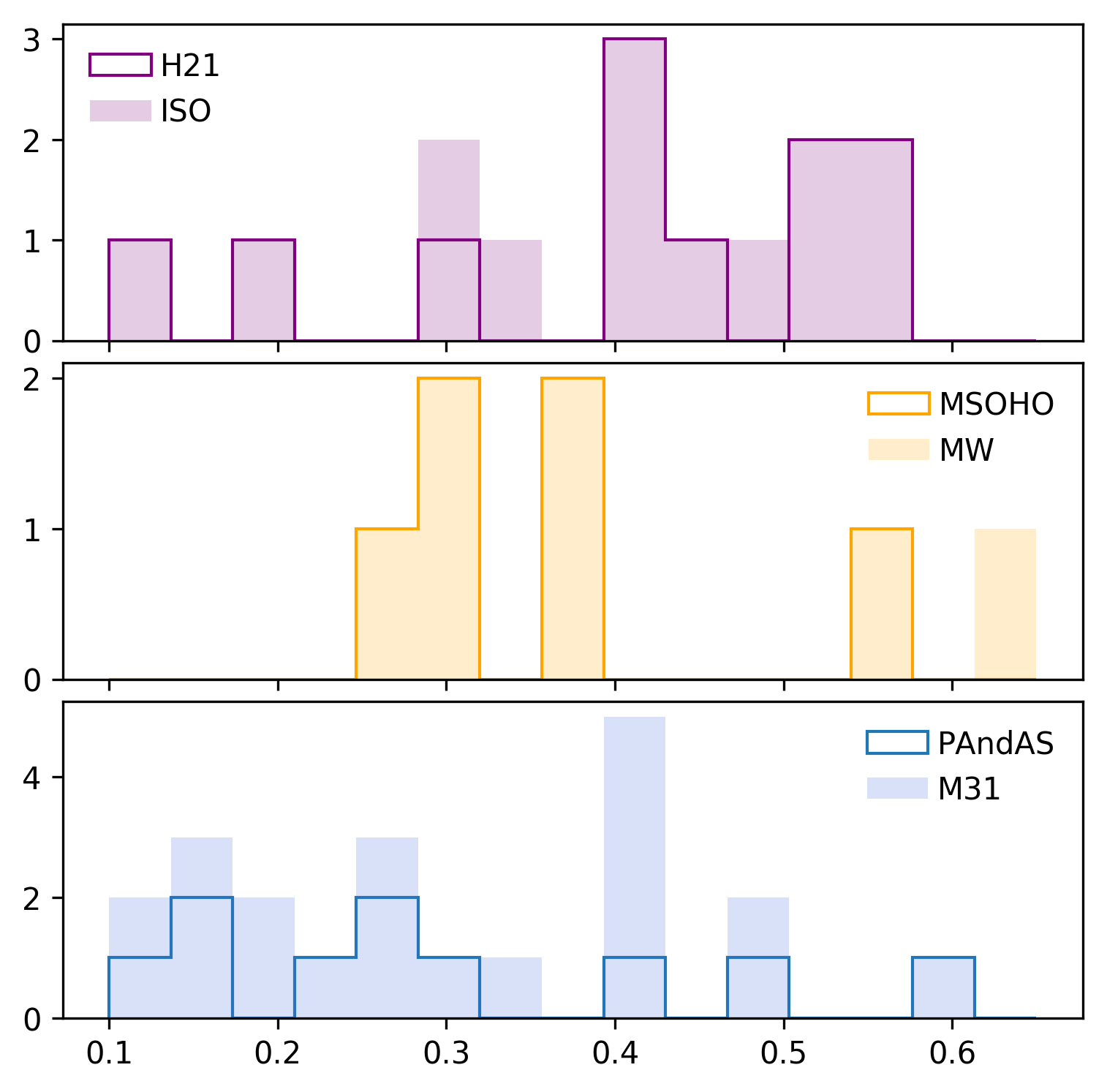} }}%
    \hfill
    \subfloat[Surface brightness ($\mu_o$)]
    {{\includegraphics[width=0.45\linewidth]{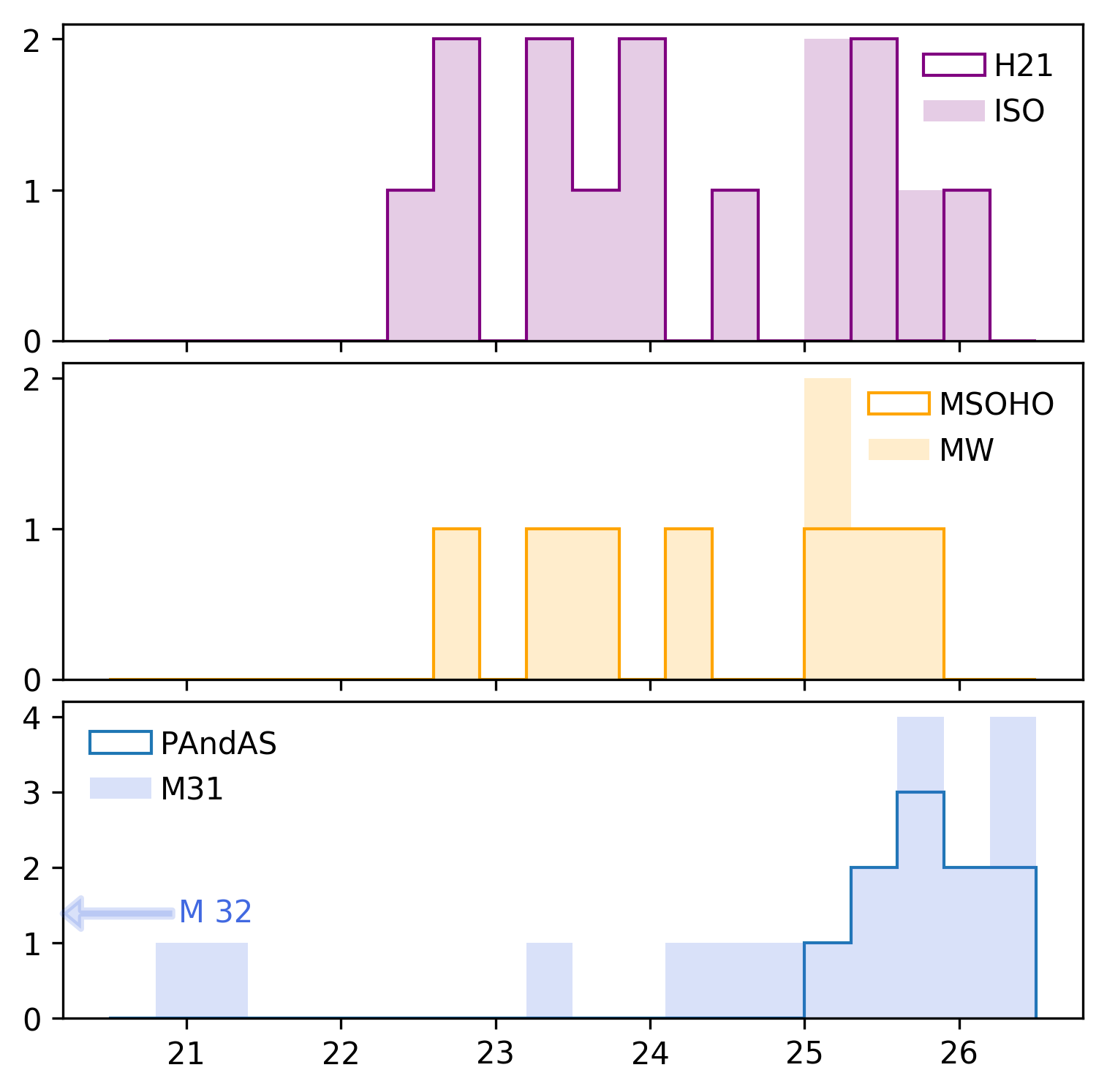} }}%
    \subfloat[V-band magnitude (M\textsubscript{v})]{{\includegraphics[width=0.45\linewidth]{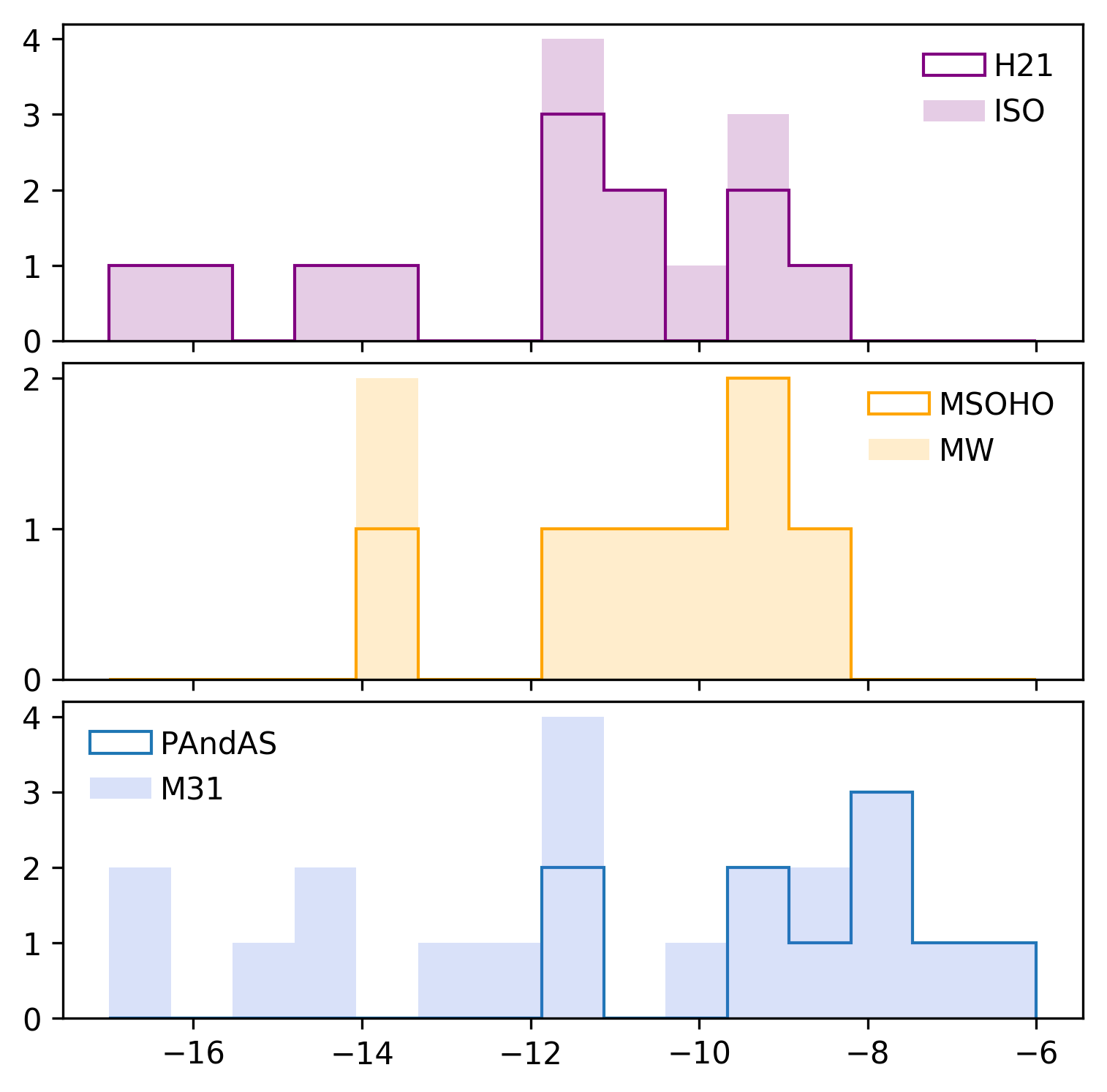} }}%
    \caption{Histograms of the distribution of structural parameters for isolated (top; purple), MW (middle; yellow) and M\,31 (bottom; blue) dwarfs. Dwarfs indicated with arrows lie far outside the range displayed, with actual values listed in Tables \ref{tab:iso}, \ref{tab:mw} and \ref{tab:M31}.}%
    \label{fig:hist_1}%
\end{figure*}

\begin{figure*}
	\includegraphics[width=\linewidth]{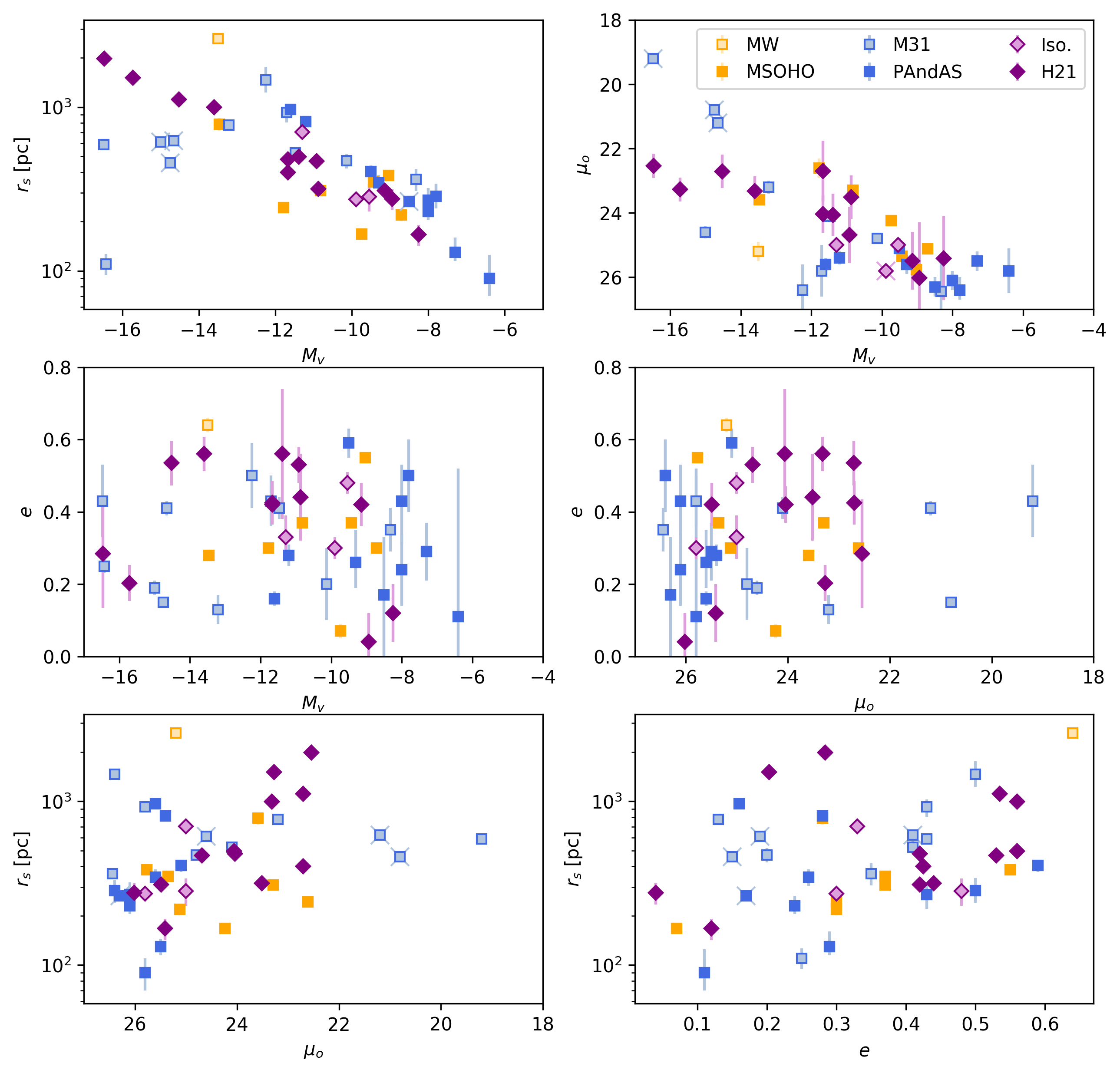}
    \caption{Kormendy relations and related trends for the observed structural parameters of the three dwarf populations. The MW population is shown in yellow, the M\,31 population is shown in blue and the isolated population is shown in purple. The darker subsets are the uniform surveys (MSOHO, PAndAS and H21 respectively). Points with no error bars are indicated with an X.} 
    \label{fig:panel6}
\end{figure*}

Figure~\ref{fig:hist_1} shows histograms of the distribution of dwarfs in each of the three data sets for the four main structural parameters: ellipticity ($e$), central surface brightness in the V-band ($\mu_o$), half-light radius ($r_s$), and absolute magnitude in the V-band (M\textsubscript{v}). For each of the three populations, the colors indicate the full dataset, and the outlines indicate dwarfs belonging to the three largest, homogeneous, studies (H21, MSOHO and PAndAS, for the isolated dwarfs, MW dwarfs and M\,31 dwarfs, respectively).

We attempt to quantitatively compare these histograms through the use of the two-sample Kolmogorov–Smirnov test, which allow us to determine the probability that two datasets are drawn from the same underlying distribution. We note that our sample sizes are not large for any population, and that there are some known "weird" objects in our samples (e.g., M32 has very few counterparts in the Local Universe, and none in the Local Group). Therefore, to better understand the sensitivity of the derived KS test probability values to individual measurements in our datasets, we perform multiple realisations for each comparison. Specifically, we perform the KS test two hundred times for each comparison, and for each realisation we ignore up to two galaxies in each of the two datasets considered. We then report the median and $16^{th}/84^{th}$ percentiles of the resulting probability values in Table~\ref{tab:ks1}. We also report the "standard value" (a single value derived on application of the test to the full dataset). As expected, the median value derived in this way is always within $1-\sigma$ of the standard value. We highlight in bold any values of $P\lesssim0.05$, which indicates our null hypothesis (the two datasets are drawn from the same population) can likely be rejected.

Finally, Figure~\ref{fig:panel6} shows the structural parameters discussed in this section plotted against each other. Various, well-known trends can be seen (e.g., \citealt{kormendy1985}) in each population, and we discuss these further in Section~\ref{sec:discussion}.

\begin{table*}
    \centering
\begin{tabular}{l l||cc|cc|cc|cc}
    \multicolumn{2}{c||}{Data Sets Compared} & \multicolumn{2}{c|}{$r_s$} & \multicolumn{2}{c|}{$e$} & \multicolumn{2}{c|}{$\mu_o$}& \multicolumn{2}{c}{M\textsubscript{v}}\\ 
    && P\textsubscript{full} & P\textsubscript{med} & P\textsubscript{full} & P\textsubscript{med} & P\textsubscript{full} & P\textsubscript{med} & P\textsubscript{full} & P\textsubscript{med}\\
    \hline
    \multirow{5}{*}{MW}& MSOHO & 1.00 & 1.00$^{+0.00}_{-0.05}$ & 1.00 & 1.00$^{+0.00}_{-0.02}$ & 1.00 & 1.00$^{+0.00}_{-0.01}$ & 1.00 & 1.00$^{+0.00}_{-0.05}$\\
                           & M\,31   & 0.5& 0.58$^{+0.22}_{-0.31}$ & 0.39 & 0.43$\pm$0.19 & 0.14 & 0.21$^{+0.07}_{-013.}$ & 0.45 & 0.51 $^{+0.14}_{-0.11}$\\
                           & PAndAS & 0.95 & 0.94$^{+0.05}_{-0.21}$ & 0.24 &0.24$^{+0.24}_{-0.13}$ & {\bf 0.00} & {\bf 0.01$\pm$0.01} & {\bf 0.05} & {\bf 0.06$^{+0.05}_{-0.03}$}\\
                           & Solo & 0.45 & 0.49$^{+0.27}_{-0.19}$ & 0.45 & 0.55$^{+0.22}_{-0.28}$ & 0.90& 0.85$^{+0.14}_{-0.17}$ & 0.69 & 0.69$^{+0.13}_{0.22}$ \\
                           & H21 & 0.29 & 0.36$^{+0.20}_{-0.19}$ & 0.29 & 0.40$^{+0.17}_{-0.22}$ & 0.73 & 0.73$^{+0.20}_{-0.26}$ & 0.41 & 0.47$^{+0.17}_{-0.23}$\\
    \hline
    \multirow{4}{*}{MSOHO}& M\,31   & 0.22 & 0.27$\pm$0.16 & 0.51 & 0.54$\pm$0.24& 0.22 & 0.33$^{+0.11}_{-0.16}$ & 0.51 & 0.55$^{+0.12}_{-0.08}$\\
                           & PAndAS & 0.99 & 0.94$^{+0.04}_{-0.22}$ & 0.38 & 0.51$^{+0.22}_{-0.30}$& {\bf 0.01} & {\bf 0.02$^{+0.02}_{-0.01}$} & 0.06 & {\bf 0.08$^{+0.07}_{-0.04}$}\\
                           & Solo & 0.20 & 0.27$^{+0.11}_{-0.16}$ & 0.20 & 0.28$^{+0.19}_{-0.10}$ & 1.00 & 0.98$^{+0.02}_{-0.16}$ & 0.40 & 0.47$\pm$0.26\\
                           & H21 & 0.12 & 0.17$^{+0.09}_{-0.11}$ & 0.12 & 0.19$^{+0.08}_{-0.12}$& 0.93 & 0.92$^{+0.07}_{-0.25}$ & 0.21 & 0.28$^{+0.19}_{-0.17}$\\
    \hline
    \multirow{3}{*}{M\,31}& PAndAS & 0.40 & 0.40$^{+0.20}_{-0.17}$ & 1.00 & 1.00$^{+0.00}_{-0.04}$& 0.22 & 0.25$^{+0.10}_{-007.}$ & 0.22 & 0.22$^{+0.0.07}_{-0.10}$\\
                           & Solo & 0.74 & 0.69$^{+0.08}_{-0.09}$& 0.26 & 0.26$^{+0.10}_{-0.08}$& 0.10 & 0.13$^{+0.07}_{-0.05}$ & 0.50 & 0.50$^{+011.}_{-0.08}$\\
                           & H21 & 0.49 & 0.49$^{+0.27}_{-0.10}$ & 0.17 & 0.20$^{+0.13}_{-0.09}$& 0.07 & {\bf 0.07$^{+0.07}_{-0.03}$} & 0.66 & 0.66$^{+016.}_{-0.15}$\\
    \hline
    \multirow{2}{*}{PAndAS}& Solo & 0.15 & 0.20$^{+0.15}_{-0.11}$& 0.15 & 0.20 $^{+0.13}_{-0.11}$& {\bf 0.00} & {\bf 0.00$\pm$0.00} & {\bf 0.01} & {\bf 0.02$^{+0.03}_{-0.01}$}\\
                           & H21 & 0.19 & 0.20$^{+0.20}_{-0.09}$ & 0.37 & 0.43$^{+0.24}_{-0.21}$0& {\bf 0.00} & {\bf 0.00$\pm$0.00} & {\bf 0.05} & {\bf 0.05$^{+0.06}_{-0.02}$}\\
    \hline
    \multirow{1}{*}{Solo}& H21 & 1.00 & 1.00$^{+0.00}_{-0.02}$ & 1.00 & 1.00$\pm$0.00 & 1.00 & 1.00$^{+0.00}_{-0.11}$ & 1.00 & 1.00$^{+0.00}_{-0.02}$\\
    \hline
        \hline
    MW \& M\,31 & Solo & 0.54 & 0.59$^{+0.09}_{-0.10}$ & 0.20 & 0.24$^{+0.13}_{-0.10}$ &0.22 & 0.24$^{+0.13}_{-0.09}$& 0.70 & 0.73$^{+0.10}_{-0.26}$ \\
\end{tabular}
\caption{Results of the KS tests between the structural parameter distributions for the different subsets of dwarf populations.  P\textsubscript{full} refers to the KS test preformed using the entire subsets.  P\textsubscript{med} is the median value and associated errors found using the random sampling as described in the text. Values highlighted in bold are deemed statistically significant. The final line of the table (MW \& M\,31 vs. Solo) compares all satellites to isolated dwarfs without distinguishing between the M\,31 and MW satellites.}\label{tab:ks1}
\end{table*}

\subsection{Dynamical Masses and Kinematics}\label{sec:mass}

We now consider the stellar velocity dispersions ($\sigma_s$), dynamical masses (M\textsubscript{dyn}, derived within their 3-D half-light radius, $r_{\sfrac{1}{2}}$) and mass-to-light ratios ($M/L$, again derived within $r_{\sfrac{1}{2}}$) for these dwarfs. We derive M\textsubscript{dyn} following \citep{Kirby2014}, where: 
\begin{equation}
    M_{dyn}=4 G ^{-1} \sigma_s ^2 r_s \text{ , where } \frac{4}{3}r_s \approx r_{\sfrac{1}{2}}
\end{equation}

\begin{figure*}%
    \centering
    \subfloat[Stellar velocity dispersion ($\sigma_s$, km s$^{-1}$)] {{\includegraphics[width=0.4\linewidth]{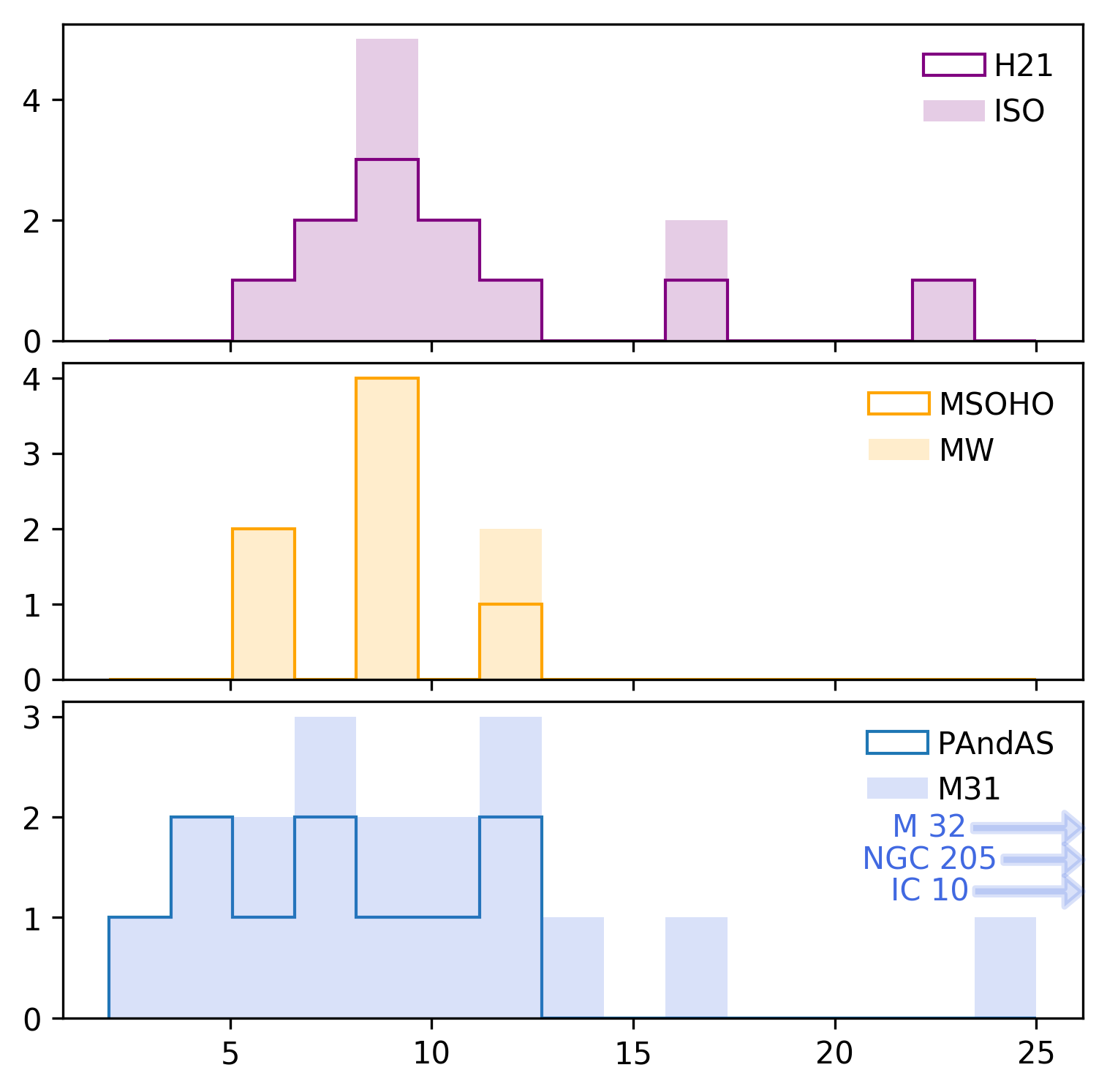} }}%
    \subfloat[Dynamical (M\textsubscript{dyn}, $10^6$ M$_{\odot}$)]{{\includegraphics[width=0.4\linewidth]{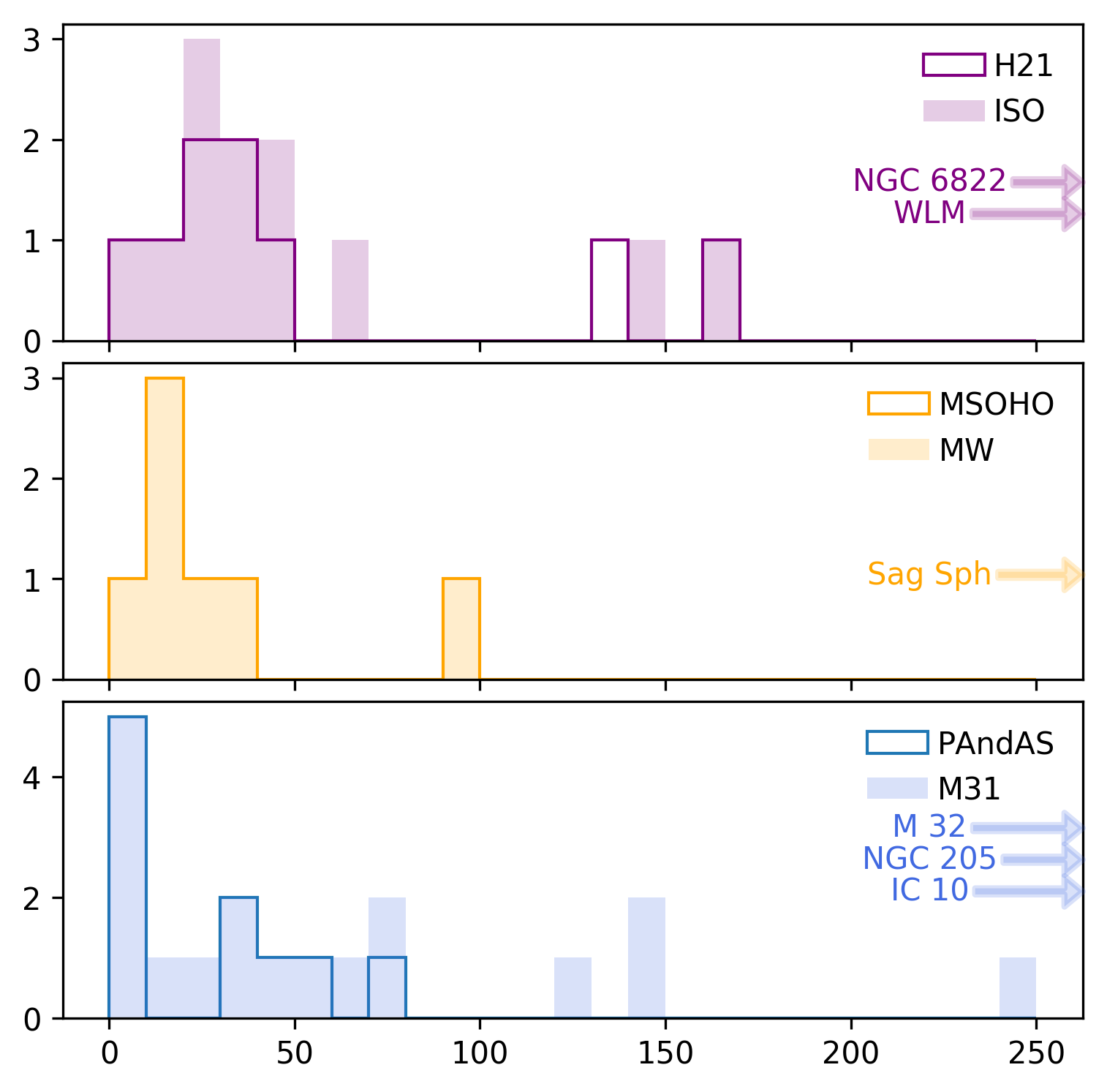} }}%
    \hfill
    \subfloat[Mass-to-light ratio (M\textsubscript{dyn}/L\textsubscript{v}, M$_{\odot}$ L$_{\odot}^{-1}$)]
    {{\includegraphics[width=0.4\linewidth]{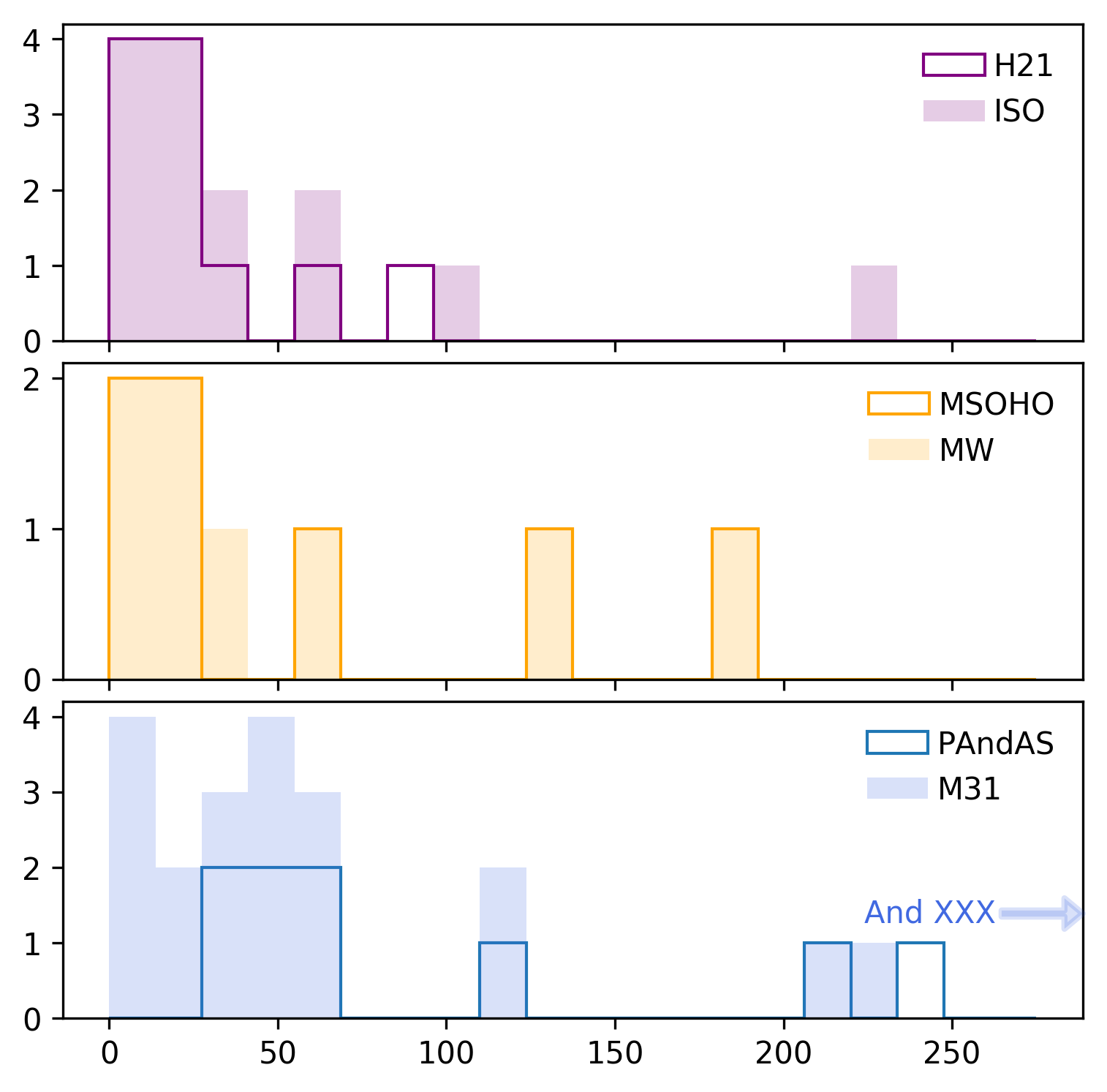} }}%
    \caption{ Histograms of the distribution of dynamical parameters for isolated (top; purple), MW (middle; yellow) and M 31 (bottom; blue) dwarfs. Dwarfs indicated with arrows lie far outside the range displayed, with actual values listed in Tables 1, 2 and 3}%
    \label{fig:hist_2}%
\end{figure*}

\begin{figure*}
	\includegraphics[width=\linewidth]{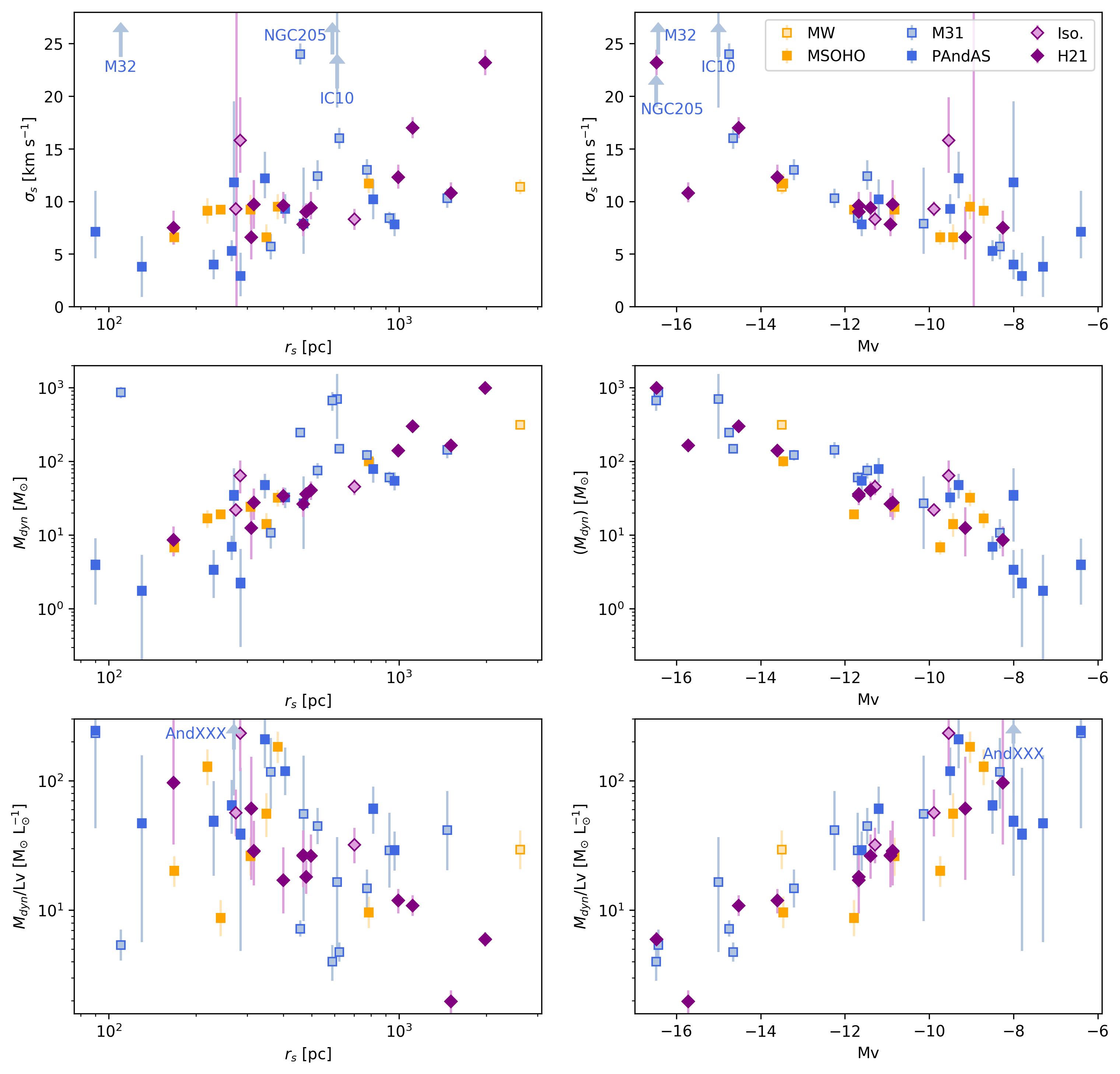}
    \caption{Relationships between the dynamical parameters (specifically, velocity dispersions, dynamical masses, and mass-to-light ratios) and key structural parameters (specifically, half-light radius and absolute magnitudes) for the dwarf galaxy populations.}
    \label{fig:dyn6}
\end{figure*}

\noindent The resulting values of M\textsubscript{dyn} and $M/L$ are listed in Tables~\ref{tab:iso}, \ref{tab:mw} and \ref{tab:M31}. Histograms are shown in Figure~\ref{fig:hist_2} and follow the style of Figure~\ref{fig:hist_1}. We perform the same KS tests as described in Section~\ref{sec:structure} and tabulate the derived probabilities in Table~\ref{tab:ks2}. Figure~\ref{fig:dyn6} shows these dynamical parameters as a function of scale radius and luminosity.

It is worth noting here that the newly derived dynamical masses for all the galaxies considered as dwarf irregular or transition-type studied by H21, use radial velocity dispersions and "half-light" radii measured from the same stellar population (i.e, evolved giants, mostly red giant branch stars). Previously, for example in \cite{Kirby2014}, the radial velocity dispersion based on individual spectroscopic measurements of evolved giants are sometimes (e.g., for IC\,1613, NGC\,6822) necessarily combined with the half-light radius based on integrated light measurements of the same galaxies. Since dIrrs usually have a prominent young, bright, blue centrally concentrated population, the "half-light" radius that is measured by integrated light measurements does not correspond to the "half-mass radius of the relevant stellar tracer" that is formally required by Equation~1. At least for IC\,1613 and NGC\,6822, this was previously unavoidable since the relevant measurements did not exist for these galaxies until the analysis by H21.

\begin{table*}
    \centering
\begin{tabular}{l l||cc|cc|cc}
    \multicolumn{2}{c||}{Data Sets Compared} & \multicolumn{2}{c|}{$\sigma_s$} & \multicolumn{2}{c|}{M\textsubscript{dyn}} & \multicolumn{2}{c}{M\textsubscript{dyn}/L\textsubscript{v}}\\ 
    && P\textsubscript{full} & P\textsubscript{med} & P\textsubscript{full} & P\textsubscript{med} & P\textsubscript{full} & P\textsubscript{med} \\ %$^{+0.}_{-0.}$
    \hline
    \multirow{5}{*}{MW}& MSOHO & 1.00 & 1.00$^{+0.00}_{-0.03}$ & 0.97 & 0.95$^{+0.05}_{-0.25}$ & 1.00 & 0.99$^{+0.01}_{-0.09}$ \\
                           & M\,31   & 0.18 & 0.22$^{+0.0.07}_{-0.05}$ & 0.20 & 0.25$^{+0.16}_{-0.13}$ & 0.83 & 0.84$^{+0.13}_{-0.27}$ \\
                           & PAndAS & 0.38 & 0.37$^{+0.27}_{-0.09}$ & 0.38 & 0.38$^{+0.18}_{-0.10}$ &  0.11 & 0.14$^{+0.14}_{-0.09}$ \\
                           & Solo & 0.52 & 0.70$^{+0.12}_{-0.18}$ & 0.23 & 0.27$^{+0.29}_{-0.14}$ & 1.00 & 0.96$^{+0.04}_{-0.17}$  \\
                           & H21 & 0.57 & 0.72$^{+0.12}_{-0.20}$ & 0.19 & 0.29$^{+0.23}_{-0.14}$ & 0.88 & 0.67$^{+0.15}_{-0.28}$ \\
    \hline
    \multirow{4}{*}{MSOHO}& M\,31   & 0.22 & 0.22$^{+0.12}_{-0.08}$ &  0.07 &  0.12$^{+0.07}_{-0.06}$ & 0.71 & 0.77$^{+0.17}_{-0.36}$ \\
                           & PAndAS & 0.42 & 0.47 $^{+0.21}_{-0.16}$ & 0.42 & 0.36$^{+0.17}_{-0.15}$& 0.08 & 0.11$^{+0.12}_{-0.07}$\\
                           & Solo & 0.37 & 0.52$^{+0.24}_{-0.30}$ & 0.10 & 0.14$^{+0.15}_{-0.08}$ & 1.00 & 0.92$^{+0.05}_{-0.15}$ \\
                           & H21 & 0.27 & 0.40 $^{+0.30}_{-0.14}$ & 0.09 & 0.12$^{+0.17}_{-0.06}$& 0.97 & 0.70$^{+0.20}_{-0.16}$ \\
    \hline
    \multirow{3}{*}{M\,31}& PAndAS & 0.36 & 0.36$^{+0.13}_{-0.06}$ & 0.22 & 0.21$\pm$0.09 & 0.24 & 0.27$^{+0.13}_{-0.14}$ \\
                           & Solo & 0.64 & 0.66 $^{+0.16}_{-0.05}$ & 0.64 & 0.64$^{+0.05}_{-0.15}$ & 0.29 & 0.23$^{+0.11}_{-0.10}$ \\
                           & H21 & 0.72 & 0.73$^{+0.04}_{-0.09}$ 0.10 & 0.72 & 0.72$^{+0.06}_{-0.10}$ & {\bf 0.05} & {\bf 0.03$\pm$0.02} \\
    \hline
    \multirow{2}{*}{PAndAS}& Solo & 0.15 & 0.19$^{+0.15}_{-0.09}$ & 0.07 & 0.08$^{+0.10}_{-0.03}$& {\bf 0.01} & {\bf 0.01$\pm 0.01$}\\
                           & H21 & 0.23 & 0.26$^{+0.21}_{-0.10}$ & 0.09 & 0.13$^{+0.10}_{-0.06}$ & {\bf 0.00} &  {\bf 0.00$\pm$0.00}\\
    \hline
    \multirow{1}{*}{Solo}& H21 & 1.00 & 1.00 $\pm$ 0.00 & 1.00 & 1.00$^{+0.00}_{-0.03}$ & 0.96 & 0.92$^{+0.07}_{-0.21}$ \\
    \hline
    \hline
    MW \& M\,31 & Solo & 0.73 & 0.78$^{+0.11}_{-0.05}$ & 0.23 & 0.26$^{+0.10}_{-0.06}$ &0.48 & 0.34$^{+0.18}_{-0.16}$\\

\end{tabular}
\caption{As Table~\ref{tab:ks1}, but for the dynamical parameters.}\label{tab:ks2}
\end{table*}

\subsection{Trends with distance from the nearest massive galaxy}

\begin{figure*}
	\includegraphics[width=\linewidth]{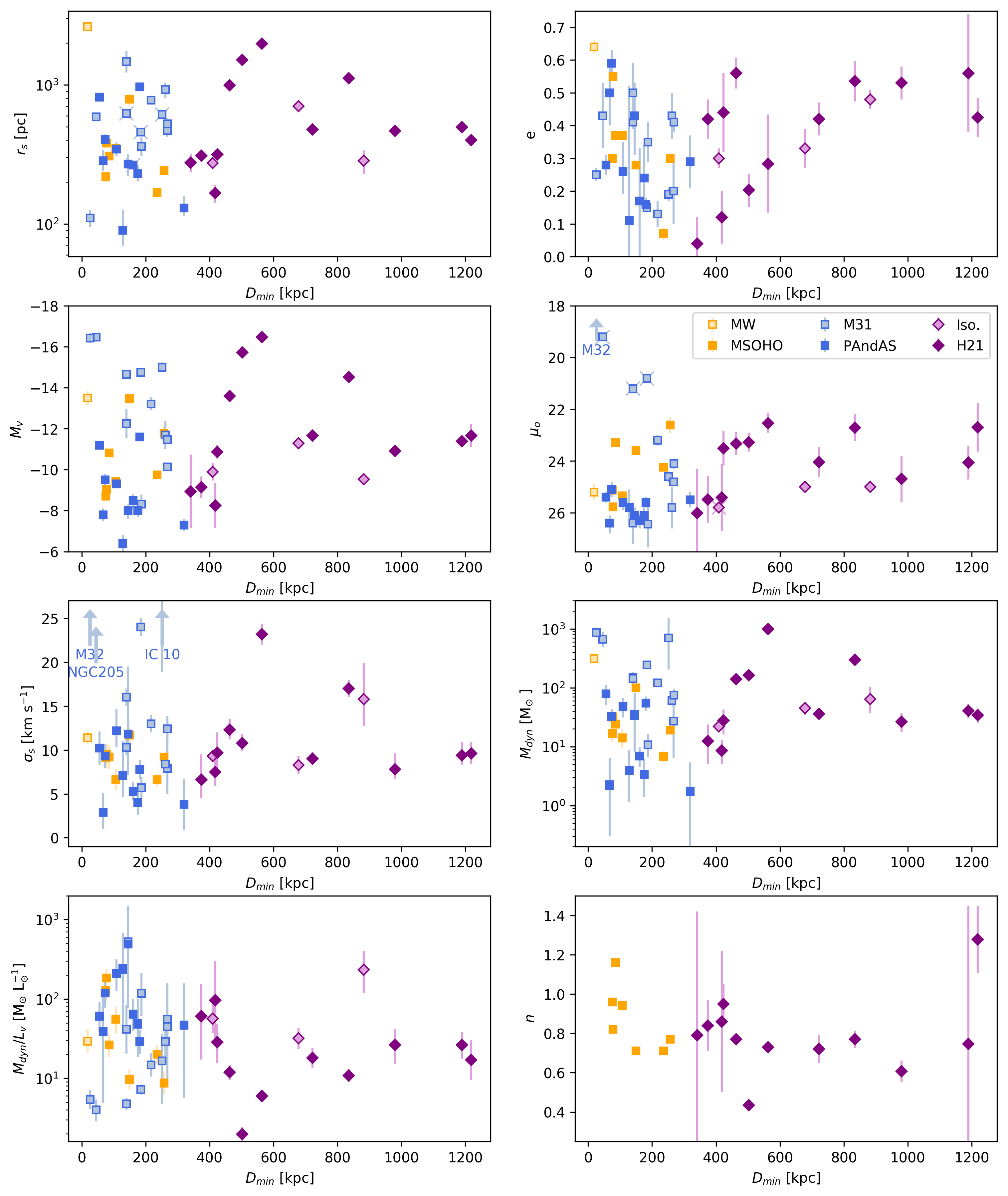}
    \caption{Trends of half-light radius ($r_s$), ellipicity ($e$), absolute magnitude (M\textsubscript{v}), surface brightness ($\mu_o$),  stellar velocity dispersion ($\sigma_s$), dynamical mass (M\textsubscript{dyn}), dynamical mass to light ratio ($M/L$) and Sersic index ($n$) with distance to the nearest massive galaxy, $D_{min}$.} 
    \label{fig:dist6}
\end{figure*}

The focus thus far has been on comparing the satellite population with the isolated population of Local Group dwarf galaxies. However, we also explore the observed structural and dynamical parameters of these galaxies as a function of their "isolation". Obviously, the only massive galaxies in the Local Group relevant to this discussion are the MW and M\,31 galaxies. They are broadly the same mass (e.g., \citealt{Eadie2019,Patel2017,BlandHawthorn2016, Penarrubia2014}), and so it is most convenient to examine the dwarf galaxy properties as a function of their distance from these systems, as a proxy for isolation. We determine $D_{min}$ for each dwarf as min(D\textsubscript{MW}, D\textsubscript{M31}), with distances to the MW and M\,31 respectively, adopting a distance modulus of 24.46 \citep{deGrijs2014} for M\,31. Figure \ref{fig:dist6} shows $r_s$, $e$, $\mu_o$, M\textsubscript{v}, $\sigma_s$, M\textsubscript{dyn}, $M/L$ and the Sersic index $n$ as a function of $D_{min}$.

An alternative metric for isolation that we considered is the tidal index (see \citealt{Karachentsev2013} for a detailed explanation). The tidal index considers the separation between a dwarf and a perturber, and incorporates the perturber's mass. We calculated the tidal index between all galaxies in our sample, including also the MW, M31, the Large and Small Magellanic Clouds, and M\,33 as possible perturbers. However, for all dwarfs, the largest tidal index is due either to the MW or M\,31, with all other galaxies much less significant, by an order of magnitude or more. This result suggests that the MW or M\,31 is the dominant perturber for the population and so D\textsubscript{min} is a good parameter to examine. 

\section{Discussion}
\label{sec:discussion}
\subsection{Comparing the satellite population of the Milky Way and M\,31}

Before we compare the satellite population of M\,31 and the MW to the isolated Local Group dwarfs, we examine if there exist major differences in the structural and dynamical parameters of the two satellite populations. There are a few notable differences between the two populations in terms of overall morphology, specifically that there are no compact or dwarf elliptical analogs to M\,32, NGC\,205,  NGC\,147 and NGC\,185 around the MW as found around M\,31. At the higher luminosity end, M\,31 has M\,33 as a companion, whereas the MW has the Magellanic Clouds. While the Magellanic Clouds account for the entire gas-rich population of MW satellites, M\,31 additionally has a dwarf starburst satellite (IC\,10) in addition to a transition-type satellite (LGS\,3). Both M\,31 and the MW appear to have planes of satellites \citep{Pawlowski2018}.

We first examine if there exist any notable differences in the structural and dynamical parameters between the galaxies with parameters derived from MSOHO and PAndAS compared to the full MW and M\,31 populations, respectively. Examination of Figures~\ref{fig:hist_1} and \ref{fig:hist_2} and the KS results in Tables~\ref{tab:ks1} and \ref{tab:ks2} show that the MSOHO sample is statistically indistinguishable from the full set of MW satellites, but this is completely unsurprising since there is only 1 additional MW satellite compared to the MSOHO sample. For the M\,31 population, Figure~\ref{fig:hist_1} shows that the PAndAS satellites are generally low luminosity/surface brightness compared to the rest of the M\,31 population. Indeed, this is not unexpected: \cite{Martin2017} specifically target the dwarf spheroidal population within the PAndAS footprint. As discussed above, this means that none of the (bright) dwarf elliptical satellites are included. Further, the PAndAS sample misses the gas rich satellites which are located outside of the PAndAS footprint. As such, we do not further concern ourselves with differences between the PAndAS-studied satellites and the rest of the M\,31 population, since the PAndAS-studied sample was never expected to be, nor described as, a representative subset of the diverse M\,31 satellite population.

Comparison of the population of the MW and M\,31 satellites to each other, both in Figure~\ref{fig:hist_1} and \ref{fig:hist_2} and more quantitatively with the KS results in Table~\ref{tab:ks1} and \ref{tab:ks2}, reveals that the two populations have statistically indistinguishable structural and dynamical properties across all parameters that we examine. It has previously been noted by \cite{McConnachie2006} and others that M\,31 appears to have satellites with considerably larger half-light radii than is found around the MW. Examination of the first panel of Figure~\ref{fig:hist_1} shows that there is indeed an "excess" of satellites around M\,31 with scale radii greater than around 400\,pc compared to the MW, although the latter does have 2 satellites like this (Fornax and the tidally-disrupting Sagittarius dSph). However, these differences do not rise to the level of statistical significance, and we conclude that  the MW and M\,31 satellites can be considered as a single grouping in our following comparison of satellites and isolated systems.

Prior to this discussion, however, we want to highlight one notable difference between the M\,31 and MW satellites. Namely, there are only eight MW satellites used in our analysis, and yet there are twenty one analogous systems around M\,31. Obscuration by the Galactic plane is clearly an issue for the MW satellites. However, the Galactic plane impacts dwarf detection in, at the most extreme, half the sky (between $-30 ^{\circ} < b +30 ^{\circ}$). Assuming that satellites are uniformly distributed, this obscuration can account for, at most, a factor of 2 difference between the MW and M\,31, whereas we observe a factor of 3 difference.

This difference aligns with other observations of M\,31, that together suggest a more active and rich accretion history for this galaxy in comparison to the MW. M\,31 is observed to have more halo globular clusters (113 outer halo globular clusters in M\,31 compared to 19 in the MW; \citealt{Wang2019}), and many more globular clusters are associated with the disk of M\,31 compared to the MW. \cite{Hammer2007} compared the integrated properties of the MW and M\,31 to other galaxies and concluded that the MW had undergone a relatively passive evolution whereas that of M\,31 was more "typical". Indeed, with respect to their satellite populations, \cite{Wang2021} recently compared the MW's satellite luminosity function to other MW-like galaxies and showed that the MW is somewhat unusual insofar as it lacks satellites (considering dwarfs with $M_v>-12$). The Saga Survey \citep{Geha2017, Mao2020} also puts the MW and M\,31 satellite systems into a broader context, and show that while the MW has a dearth of satellites as compared to M\,31, they both lie within the scatter of MW-like analogues. Thus, the difference in the number of brighter satellites observed here aligns with this overall picture of M\,31 as galaxy with a more active history than the MW. 

\subsection{Examining the trends of the structural and dynamical parameters}

Across all structural and dynamical parameters that we study, neither visual inspection of the histograms in Figures~\ref{fig:hist_1} and \ref{fig:hist_2} nor the KS test results in Tables~\ref{tab:ks1} and \ref{tab:ks2} reveals any statistically significant evidence that the one-dimensional parameters distribution functions for  satellites compared to isolated systems are different.

Evidence for a more complex story emerges when we consider the Kormendy relations and related diagrams shown in Figure~\ref{fig:panel6} and \ref{fig:dyn6}. Particularly striking is the relation between $r_s$ and M\textsubscript{v} in the top-left panel of Figure \ref{fig:panel6}. Here, the isolated dwarfs (purple) define a remarkably linear relation between scale-radius and magnitude. The satellite populations (blue and yellow) follow the same trend, but with notably more scatter around the relation. Inspection of the top-right panel of the same figure shows similar behaviour for the central surface brightness. That is, while all dwarfs appear to follow the same broad trend in surface brightness with luminosity, the relation defined by the isolated systems appears more tightly defined than that for the satellite systems. 

This general behaviour is also visible in some of the derived dynamical trends, perhaps most obviously in M\textsubscript{dyn} vs. $r_s$. Indeed, the common element to all of these trends in their reliance on the physical scale of the dwarf (M\textsubscript{dyn} is calculated using $r_s$, and surface brightness is a measure of the concentration of luminosity within a scale radius). However, we note that we are not able to say that this increased scatter has very high statistical significance. We have tried to use multiple parameterizations of the scatter for these relations, and we find what we can see by eye: the scatter is generally larger for satellites than for isolated systems. In no case, however, are we able to claim this increased scatter is at a high significance. This is likely due to the reality of the Local Group of galaxies: there are simply not enough galaxies in each of our subsets to be able to claim that weak differences in populations are statistically significant. Given the known, observable sample available, only the most extreme differences between populations will ever be classed as statistically significant, although this does not mean that weak differences do not exist, nor that they are uninteresting to consider. 

We must use our judgement as to what importance to assign to this apparent increase in scatter. It is interesting to consider that this behaviour may then naturally be explained through a single process, namely the tidal influence of the massive galaxies on the scale radius of the dwarf. For the case of dark matter dominated dwarf spheroidal populations, this has been examined from a theoretical perspective by \cite{Penarrubia2008}. They find that these systems follow remarkably predictable tracks in these 2-parameter plots as mass is stripped. Initially, as the outskirts of the dwarf is stripped (which results in the loss of dark matter but little luminosity), the scale radius of the dwarf will increase as it adjusts to the reduced dark matter content of the dwarf. Then, as stripping proceeds and luminous matter is lost, the scale radius decreases significantly as the overall luminosity decreases. 
Indeed, following \cite{Errani2015}, \cite{Fattahi2018} predict the progenitors of various Local Group dwarfs (especially satellites) prior to tidal stripping, and find the progenitor population occupies a parameter space that is more tightly defined than that occupied by the present-day dwarfs. It seems reasonable to suggest, therefore, that a similar process may be at work here, and could naturally explain why the satellites appear to show some evidence for having a larger scatter in scale radius and related properties. We also note that stripping, while present, is unlikely to be dominant for all satellites, since we might then expect there to be more notable differences in the overall 1-D distributions of these parameters between the populations. 

\subsection{How isolated is isolated?}

The original definition of the {\it Solo} sample with respect to isolation is $D_{min} > 300$\,kpc from the MW or M\,31, and was based on the virial radii of the dark matter halos in which these galaxies are expected to reside. We investigate if this "boundary" does indeed represent a meaningful definition of isolation by examination of the trends with distance from the nearest massive galaxy in Figure~\ref{fig:dist6}.

Examination of the panels in Figure 6 consistently show that the identified satellites of M\,31 and the Milky Way have generally large spreads on the y-axis of each panel, and that these spreads are much larger than for those dwarfs that have large values of $D_{min}$ (this is true to varying extent for every panel except the last). There appears to be a transition area around $D_{min} \simeq 300 - 400$\,kpc, to the left of which dwarf galaxies occupy a broad range of parameter space, and to the right of which the range that is occupied is considerably less. This is consistent with our earlier interpretation of the effect of tides, but what is interesting here is that the effect does not match exactly our original "threshold" for isolation of $D_{min} = 300$\,kpc. Rather, the parameters of the five isolated galaxies between $D_{min} \simeq 300 - 400$\,kpc (Phoenix, Leo T, Perseus, And XVIII and And XXVIII) appear to bear more similarity as a group to the satellite population than the more isolated galaxies; this is most apparent when examining their half-light radius, their magnitudes, and their surface brightnesses.

In addition to these parameters, another feature present in Figure~\ref{fig:dist6} is the dearth of low ellipticity dwarfs at large distances visible in the top-right panel. As we get close to the more massive galaxies, we find dwarf galaxies with ellipticities spanning $0 \lesssim e \lesssim 0.6$. However, further than $D_{min} \sim 500$\,kpc, no low ellipticity systems are found, and indeed all very isolated systems approximately 1\,Mpc from either the MW or M\,31 have a broadly similar ellipticity of around $e \sim 0.5$. We also emphasise that the ellipticity that we measure is the ellipticity of the older stellar populations in all the dwarfs i.e., for the isolated systems we are not looking at recently formed stars in a young disk, as might be the case if we used integrated light measurements for this comparison.

We conclude from Figure~\ref{fig:dist6} that, across all panels, the differences between the "isolated" galaxies and the "satellite" galaxies are maximised if we consider a distance threshold of $D_{min} \simeq 400$\,kpc, not 300\,kpc. The transition distance at which this occurs is occupied by a group of five dwarfs (Phoenix, Leo T, Perseus, And XVIII and And XXVIII). 
%added paragraph \cite{Gill2005}
From simulations, the distance range between $1 - 2$ times the virial radius of the host is populated by many dwarfs which have previously undergone an interaction with the host. Indeed, \cite{Gill2005} suggests that 50\% of dwarfs within this distance range have previously interacted with the host. For M31 and the MW, this distance range corresponds to $D_{min} \sim 300-600$\,kpc. In this range, there are 8 isolated dwarfs in our sample, with possibly 5 looking more similar to satellites than more distant dwarfs. Qualitatively, these findings would seem to align with expectations for backsplash systems.

\subsection{Building a consistent interpretation}

While we do not see any clearly statistically significant differences between satellite and isolated galaxies in any of the parameter space we examine, the items discussed above allow for a consistent interpretation of the data at hand. Specifically:

\begin{itemize}
    \item Most dwarfs within approximately 400\,kpc of the MW and M\,31 are satellites or backsplash galaxies. This is consistent with studies of these populations in simulations. \cite{Teyssier2012} suggests that 13\% of systems between $300 - 1500$\,kpc have actually orbited around the host galaxy. For the Local Group, they suggest that Cetus, Tucana, NGC 6822,  Phoenix and Leo T have passed through the Milky Way. This would constitute one-third of our ``isolated'' sample. A similar study by \cite{Buck2019} reaches similar conclusions regarding NGC~6822 and Leo T, but they find Cetus and Phoenix are slightly less likely to be backsplash systems (with probabilities of $P \simeq 0.36$ and 0.43, respectively; Tucana was not included in their analysis). Further, \cite{Blana2020} also suggests that Leo T is a backsplash dwarf, while Phoenix is argued to be on its first in-fall. Actual measurements of the proper motion of NGC~6822 in Paper III, however, suggest that NGC~6822 is not a backsplash system. These results highlight the complex and uncertain nature of identifying backsplash galaxies without three dimensional velocity information.
    
    \item Interactions of the population of satellite galaxies with their hosts generally causes an increase in the scatter of the underlying trends between the structural and dynamical parameters discussed. This is especially true with respect to the scale radius of the dwarf (e.g., top-left panel of Figure~\ref{fig:panel6}), as well as surface brightness (compare the scatter in surface brightness for galaxies interior and exterior to $D_{min} \sim 400$\,kpc in Figure~\ref{fig:dist6});
    \item Satellite galaxies have a broader range of ellipticities than isolated systems, and in particular satellite populations have more circular satellites than are found in the field. This is consistent with interactions in the satellite environment making systems more spheroidal, for example as proposed by \cite{Kazantzidis2011}, and references therein. These authors invoke a suite of physical processes to "tidally stir" disk-like populations to become spheroidal-like. Interestingly, \cite{Lokas2012} simulate the evolution of various observed parameters for dwarf galaxies (M\textsubscript{v}, $\mu_o, e$) as the system is tidally stirred. During their interactions, the ellipticity of the dwarfs are scattered, and are not systematically increased or decreased. Given many of these systems start off with relatively high $e$, the result is the production of many dwarfs with smaller $e$, consistent with our findings. But we also note that the assumption that the stars start off in a cold disk is not necessarily correct (e.g., \citealt{Kaufmann2007}), and this is supported by the fact that the old populations of several isolated dwarf galaxies are not seen to possess any significant rotation (\citealt{Kirby2017}). 
    \item Given that our interpretation of these plots puts heavy weight on tidal effects, it is also worth noting how the dynamical parameters vary with distance from host. There are only 5 galaxies in our sample that have velocity dispersions $\sigma_v \lesssim 6$\,km\,s$^{-1}$. All of them are satellites. Further, examination of the derived dynamical masses for all the dwarfs (which is a function of both $\sigma_v$ and $r_s$) shows even more clearly that the galaxies with the smallest dynamical masses are satellites. Again, this is broadly consistent with the idea that the satellite populations contain systems that have been notably affected by tidal stripping.
\end{itemize}

A consistent picture can therefore emerge from these comparisons, that is also consistent with our general understanding of the likely dominant mechanisms at work for satellites, i.e., tides. The absence of circular galaxies in the field is particularly striking. While it is nearly certainly the case that some highly elliptical satellite galaxies are tidally disrupting (e.g., Hercules; see \citealt{Deason2012,Roderick2015,Garling2018} among others), it is also clear that many satellites can be intrinsically elliptical, without the need for tides. Our findings suggest that more spheroidal satellites, on the other hand, would not be nearly spherical were it not for their proximity to a massive host.

\subsection{Revisiting completeness of the sample}

As discussed in Section 2, the sample on which our analysis and discussion is based is limited in both magnitude and surface brightness, in an attempt to remove the difficulties caused by inhomogeneous detection limits for the three populations. Would our conclusions change significantly if somehow completeness was still an issue, especially for the isolated subset of dwarfs? 

With respect to the behaviour of ellipticity, it seems difficult to understand how this would change, since there is no clear reason why nearly circular dwarfs at large distances from either the MW or M\,31 would be harder to detect than elliptical ones. Indeed, examination of Figure~\ref{fig:panel6} shows there is no trend of ellipticity with either $M_v$ or $\mu_o$ (the two parameters on which completeness most likely depends). Completeness would also not change our discussion about the increase in scatter of various parameters, especially $r_s$, since finding more dwarfs at fainter magnitudes or surface brightnesses will only add new points at the faint end, not change the scatter for any of the systems we already know about. However, regarding Figure~\ref{fig:dist6}, it could change our ability to discern a distance threshold that distinguishes satellites from field. If there are many systems as faint as Phoenix, Leo T, Perseus, And XVIII and And XXVIII at greater distances from the MW and M\,31, then they will likely also have smaller scale radii, lower radial velocity dispersions, and lower dynamical masses (as per the correlations in Figures~\ref{fig:panel6} and \ref{fig:dyn6}). This unseen population could essentially remove the distinguishing features that cause us to conclude that most systems are in fact satellites within D\textsubscript{min} $\simeq 400$\,kpc. We note, however, that the differences in ellipticities would still persist.

\cite{Fattahi2020} argue that, from comparison of the Local Group to cosmological simulations, there may be as many as 50 more dwarfs at least as massive as Draco currently unobserved in the field out to a distance of 3\,Mpc (a considerably larger volume than the approximately 1\,Mpc distance threshold we use in the current study). However, if we also adopt the best estimates we have for the completeness of the current searches of Local Group satellites, then these systems would mostly have to be very low surface brightness, physically extended, dwarf galaxies like Antlia 2, Crater 2 or Andromeda XIX, and occupy a quite distinct region of parameter space compared to those galaxies studied here. The coming few years, aided by the survey power of the Vera C. Rubin Observatory, the Nancy Grace Roman space telescope, and the Euclid mission, should provide us with the definitive surveys to push searches of dwarf  galaxies to low surface brightness throughout the Local Group.

\section{Summary and Conclusion}

We have used the {\it Solo} ({\it So}litary {\it Lo}cal) Dwarf Galaxy Survey (Paper I, II) to compare and contrast isolated dwarfs in the Local Group with those in close proximity to the Milky Way  and M\,31. In particular, our comparison focuses on the global structural properties of these galaxies as traced by their oldest stellar populations. In addition to their sizes, shapes, luminosities and surface brightnesses, we also examine their global dynamical properties, including their velocity dispersions, implied dark matter masses, and mass-to-light ratios. We attempt to account for different selection effects between our samples by adopting firm faint-end limits in magnitude and surface brightness. Our main findings are:
\begin{itemize}
    \item The structural and dynamical properties of the satellite populations of the MW and M\,31 are not obviously statistically different. However, it is notable that there are a lot more satellites around M\,31 than around the MW down to an  absolute magnitude $M_v \le -8$ and $\mu_o < 26.5$\,mags\,arcsec$^{-2}$ (21 around M\,31 compared to 8 around the MW (considering those satellites fainter than the Magellanic Clouds).
    \item The proximity of a massive host does not induce any statistically significant offsets between the observed structural or dynamical one-dimensional parameter distributions for satellites compared to isolated systems.
    \item Dwarfs in close proximity to a massive galaxy generally show more scatter in their Kormendy relations than those in isolation. Specifically, isolated Local Group dwarf galaxies show a much tighter trend of half-light radius versus magnitude than the satellite population, and similar effects are also seen for related and derived parameters (i.e., surface brightness, dynamical mass, and mass-to-light ratio).
    \item There is an absence of spherical (i.e., $e \lesssim 0.3$) dwarf galaxies far from large galaxies;
    \item There appears to be a transition in the structural and dynamical properties of the dwarf galaxy population at around $D_{min} \sim 400$\,kpc from the MW and M\,31, such that the dwarf population closer than this occupy a broader range of parameter space compared to dwarfs further away, and that the smallest, faintest, most circular systems are found as satellites.
    \item Differential selection effects between the surveys that have found MW, M\,31 and Local Group dwarf galaxies are important to consider, but are unlikely to significantly affect the conclusions of this current study unless our current understanding of the limits of these surveys are significantly flawed.
\end{itemize}

The limited number of systems that pass our magnitude and surface brightness cuts means that the statistical significance of any one of these findings is not high. However, together they paint a compelling picture that we interpret as pointing to the significance of tidal interactions on the population of systems within approximately 400\,kpc from the MW and M\,31. These interactions act to increase the scatter in the various relationships we observe, including potentially making some systems more spherical than they might otherwise become if left to nature rather than nurture.

\section{Data Availability}

The raw data on which this analysis is based are publicly available in the CFHT archive (accessed via the Canadian Astronomical Data Center at \url{http://www.cadc-ccda.hia-iha.nrc-cnrc.gc.ca/en/cfht/}).
Processed data may be shared on reasonable request to the corresponding author.

%%%%%%%%%%%%%%%%%%%% REFERENCES %%%%%%%%%%%%%%%%%%
\section*{Acknowledgements}

We thank Julio Navarro for some useful discussions and feedback during the preparation of this manuscript. We thank Geraint Lewis for his helpful suggestions. 
We acknowledge and respect the \textipa{l@\'k\super w@N@n} peoples on whose traditional territory the University of Victoria, where this work was conducted, stands and the Songhees, Esquimalt and \b{W}S\'{A}NE\'{C} peoples whose historical relationships with the land continue to this day. As we explore the shared sky, we acknowledge our responsibilities to honour their enduring relationships with these lands.

\bibliographystyle{mnras}
%\{chicago}
\bibliography{bib_ALL}

%%%%%%%%%%%%%%%%% APPENDICES %%%%%%%%%%%%%%%%%%%%%

%%%%%%%%%%%%%%%%%%%%%%%%%%%%%%%%%%%%%%%%%%%%%%%%%%
%%%%%%%%%%%%%%%%%%%%%%%%%%%%%%%%%%%%%%%%%%%%%%%%%%

% Don't change these lines
\bsp	% typesetting comment
\label{lastpage}
\end{document}